\documentclass[twocolumn,superscriptaddress,prl,amsmath,amstex,amssymb,citeautoscript]{revtex4-1}
\pdfoutput=1
\usepackage[english]{babel}
\usepackage{letltxmacro}
\usepackage{latexsym}
\usepackage[utf8]{inputenc}
\LetLtxMacro{\ORIGselectlanguage}{\selectlanguage}
\makeatletter
\DeclareRobustCommand{\selectlanguage}[1]{%
  \@ifundefined{alias@\string#1}
    {\ORIGselectlanguage{#1}}
    {\begingroup\edef\x{\endgroup
       \noexpand\ORIGselectlanguage{\@nameuse{alias@#1}}}\x}%
}
\newcommand{\definelanguagealias}[2]{%
  \@namedef{alias@#1}{#2}%
}
\makeatother
\definelanguagealias{en}{english}
\definelanguagealias{English}{english}
\usepackage{graphicx}
\usepackage{amsmath}
\usepackage{amsfonts}
\usepackage{amsthm}   % Theorem Formatting
\usepackage{amssymb}
\usepackage{bm}
\usepackage{color}
\usepackage[percent]{overpic}
\usepackage{soul} 
\usepackage{amssymb}
\usepackage{wasysym}
\usepackage{dsfont}
\usepackage{float}
\usepackage{mathtools}

\usepackage{hyperref}
\hypersetup{
    bookmarks=false,         % show bookmarks bar?
    unicode=false,          % non-Latin characters in Acrobat�s bookmarks
    pdftoolbar=false,        % show Acrobat�s toolbar?
    pdfmenubar=true,        % show Acrobat�s menu?
    pdffitwindow=false,     % window fit to page when opened
    pdfstartview={FitH},    % fits the width of the page to the window
    pdftitle={},    % title
    pdfauthor={Authors},     % author
    pdfsubject={},   % subject of the document
    pdfcreator={},   % creator of the document
    pdfproducer={}, % producer of the document
    pdfkeywords={many-body localization} {matrix elements} {disordered systems}, % list of keywords
    pdfnewwindow=true,      % links in new window
    colorlinks=true,       % false: boxed links; true: colored links
    linkcolor=black,          % color of internal links (change box color with linkbordercolor)
    citecolor=blue,        % color of links to bibliography
    filecolor=magenta,      % color of file links
    urlcolor=black           % color of external links
}

\newcommand{\be}{\begin{equation}}
\newcommand{\ee}{\end{equation}}
\newcommand{\bea}{\begin{eqnarray}}
\newcommand{\eea}{\end{eqnarray}}

\usepackage{graphicx}
\usepackage[colorinlistoftodos]{todonotes}
\usepackage{verbatim}

\newcommand{\ket}[1]{\mbox{$| #1 \rangle$}}

\usepackage{soul}

\newtheorem*{lemma*}{Lemma}

\begin{document}

\title{Quantum Error Correction in Scrambling Dynamics and\\ Measurement-Induced Phase Transition}

\author{Soonwon Choi}
\thanks{SC and YB contributed equally to this work.}
\affiliation{Department of Physics, University of California Berkeley, Berkeley, California 94720, USA}
\author{Yimu Bao}
\thanks{SC and YB contributed equally to this work.}
\affiliation{Department of Physics, University of California Berkeley, Berkeley, California 94720, USA}
\author{Xiao-Liang Qi}
\affiliation{Stanford Institute for Theoretical Physics, Stanford University, Stanford, California 94305, USA}
\author{Ehud Altman}
\affiliation{Department of Physics, University of California Berkeley, Berkeley, California 94720, USA}
\affiliation{Materials Science Division, Lawrence Berkeley National Laboratory, Berkeley, California 94720, USA}

\begin{abstract}
We analyze the dynamics of entanglement entropy in a generic quantum many-body open system from the perspective of quantum information and error corrections.
We introduce a random unitary circuit model with intermittent projective measurements, in which the degree of information scrambling by the unitary and the rate of projective measurements are independently controlled.
This model displays two stable phases, characterized by the volume-law and area-law scaling entanglement entropy in steady states.
The transition between the two phases is understood from the point of view of quantum error correction: the chaotic unitary  evolution protects quantum information from projective measurements that act as errors. 
A phase transition occurs when the rate of errors exceeds a threshold that depends on the degree of information scrambling.
We confirm these results using numerical simulations and obtain the phase diagram of our model.
Our work shows that information scrambling plays a crucial role in understanding the dynamics of entanglement in an open quantum system and relates the entanglement phase transition to changes in quantum channel capacity.
\end{abstract}

\maketitle
A generic unitary evolution of a quantum many-body system scrambles information.
Any local degrees of freedom that are initially in an unentangled state become increasingly more entangled with the rest of the system, making the information encoded in them effectively unrecoverable~\cite{DeutschETH,SrednickiETH,rigol2008thermalization}.
The scrambling dynamics~\cite{hayden2007black,sekino2008fast,shenker2014black,hosur2016chaos}, evidenced by the growth of the entanglement entropy toward an extensive value~\cite{Haah2017entanglement,von2018operator,nahum2018operator,jonay2018coarse}, underlies the rich complexity of quantum dynamics and the fact that simulating it is beyond the capability of classical computers.

In a realistic system, however, unitary dynamics is often interspersed by occasional measurements of local observables made by external observers either controlled or accidental. This process disentangles the measured degrees of freedom from the rest of the system, which may reduce the entanglement entropy.
Thus, it is natural to ask under what conditions the growth of entanglement is tamed to a point allowing efficient classical simulations of the quantum dynamics~\cite{aharonov2000quantum,VidalMPS2004}. 

This question has been addressed in a number of recent works.
In the special case of noninteracting fermions, quantum states with volume scaling entanglement (volume-law phase) are unstable to any small rate of measurements in local occupation basis, leading to steady states in which the entropy only scales with the boundary area of a region (area-law phase)~\cite{cao2019entanglement}. 
However, the corresponding behavior in generic interacting systems appears to be much more subtle and has not been fully understood.
On the one hand, Refs.~\cite{SkinnerNahum2018measure,LiFisher2018ZenoEffect,chan2019unitary} suggested that the interplay between the unitary dynamics and measurements can lead to a transition between two distinct phases: for sufficiently small measurement rates, the system remains stable in the volume-law phase, while it undergoes a transition into an area-law phase as the rate exceeds a certain critical value.
On the other hand, in its early version, Ref.~\cite{chan2019unitary} pointed out that this phase transition cannot be explained by a simple competition between rates of entanglement growth and measurements, as it would always predict the area-law phase for nonzero measurement rates.

In this Letter, we show that a central ingredient for understanding the entanglement phase transition is the effective quantum error correction enabled by scrambling unitary dynamics.
Using simple concepts from quantum information theory, we provide new insight on the mechanism that drives the phase transition. 
Naively, the phase transition seems to hinge on the competition between the rate of entanglement generation by unitary gates and that of disentanglement by measurements.
If this perspective is true, the volume-law phase is unstable against an arbitrarily small rate of measurements since the competition is fundamentally not symmetric.
Given a bipartition, a local unitary gate may change the entanglement only when it acts nontrivially across the boundary of two subsystems.
In contrast, the effect of the measurements could be nonlocal: by disentangling all of the measured qubits inside a subsystem, the rate of entanglement reduction may be extensive.
Thus, measurements would always overwhelm the entanglement generation and destabilize the volume-law phase.
Here, we argue that this is not the case when information scrambling is taken into account.

\begin{figure}[t!]
	\centering
	\includegraphics[width=\linewidth]{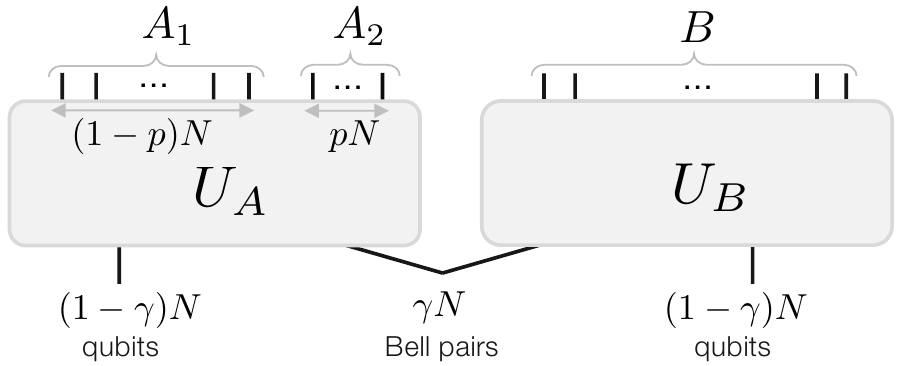}
	\caption{
		Quantum state of $2N$ qubits generated by applying unitaries $U_{A(B)}$ to $\gamma N$ Bell pairs.
		Measuring $p$ fraction of qubits ($A_2$) does not reduce the entanglement between $A$\,$=$\,$A_1A_2$ and $B$ as long as $1-p$\,$>$\,$\gamma$ in the limit $N$\,$\rightarrow$\,$\infty$.
	}
	\label{fig:toy}
\end{figure}

Our key observation is that nonlocal effects of sparse measurements are greatly suppressed due to the natural quantum error correction (QEC) property of scrambling dynamics.
If quantum information is sufficiently scrambled by unitary evolution, correlations between two subsystems are hidden in highly nonlocal degrees of freedom and cannot be revealed by any local measurements.
In such case, sparse local measurements, despite their extensive number, cannot decrease the entanglement entropy significantly.
To illustrate this point and quantify the condition under which the entanglement is robustly protected against local measurements, we improve and apply the quantum decoupling theorem to a specially constructed toy model~\cite{nielsen2002quantum,PreskillNotes}, showing the stability of the volume-law phase. 
We find that the mechanism of the protection is equivalent to that of the QEC scheme~\cite{nielsen2002quantum,schumacher2002approximate,PreskillNotes}.
Motivated by this understanding, we introduce a new model and analyze its dynamics both analytically and numerically to obtain the phase diagram.
Furthermore, we establish an exact relation between the steady-state entanglement entropy and the quantum channel capacity of quantum dynamics.

\emph{Protection against measurement}.---We now illustrate how entanglement can be protected against measurements using a well-studied toy example from quantum information theory~\cite{devetak2005private, horodecki2007quantum, abeyesinghe2009mother, PreskillNotes}.
Consider a system of $2N$ qubits ($N$\,$\gg$\,$1$) as shown in Fig.~\ref{fig:toy}.
Initially, the two halves of the system, $A$ and $B$, share $\gamma N$ Bell pairs  ($0$\,$<$\,$\gamma$\,$<$\,$1$), which control the amount of the entanglement between the two.
The two subsystems are evolved independently with unitaries $U_A$ and $U_B$, respectively. We assume $U_A$ is a random unitary drawn from the Haar distribution (or any unitary 2-design), and $U_B$ can be arbitrary.
Following this evolution, a fraction $p$ of the qubits in $A$ is measured.
The pertinent question is by how much these measurements reduce the entanglement between $A$ and $B$.
We shall show that under a certain condition the change of entanglement entropy due to the measurements vanishes in the thermodynamic limit even though an extensive number of qubits are being disentangled.
Note that this result can be generalized to incorporate measurements performed on both $A$ and $B$ by sequentially analyzing the effect of measurements.

We first simplify the problem.
Since $U_B$ does not affect the entanglement, we may replace $B$ with its minimal effective degrees of freedom $\tilde{B}$ entangled with $A$, i.e., the original $\gamma N$ entangled qubits.
Also, we divide $A$ into two parts: subsystem $A_1$ refers to the unmeasured qubits and subsystem $A_2$ contains the measured ones. 
We now apply the decoupling theorem~\cite{devetak2005private,horodecki2007quantum,abeyesinghe2009mother,PreskillNotes} to this setup, which will imply that, for a sufficiently small measurement fraction $p$, the reduced density matrix of $A_2$ and $\tilde{B}$ approximately factorizes 
\begin{align}
\label{eqn:decoupling}
\mathbb{E}_{U_A}\left[ || \rho_{A_2\tilde{B}} (U_A) - \rho_{A_2}^\textrm{max} \otimes \rho_{\tilde{B}} ||_1\right] \leq
 2^{-(1-2p-\gamma)N/2}.
\end{align} 
Here, the left-hand side denotes the distance, in the $L_1$ norm, between the exact density matrix $\rho_{A_2\tilde{B}}$ and a factorized one where $ \rho_{A_2}^\textrm{max}$ is the maximally mixed state on the measured part $A_2$.
$\mathbb{E}_U \left[ \cdot \right]$ represents averaging over the random unitaries.
The inequality implies that the measured qubits are effectively decoupled from $\tilde{B}$ for $N$\,$\gg$\,$1$, provided that the number of unmeasured qubits $A_1$ is more than half of the total system $A\tilde{B}$, or equivalently 
\begin{align}
\label{eqn:criterion}
\gamma + p <{1\over 2}\left(1+\gamma\right),
\end{align}
or simply $\gamma < 1-2p$. If this inequality is satisfied, then any observable in $A_2$ contains no information about $\tilde{B}$ and vice versa. Therefore, measuring one subsystem does not affect the other, up to an error exponentially small in $N$.
In particular, any projection (due to measurements) acting on $A_2$ does not alter $\rho_{\tilde{B}}$, and the entanglement entropy of subsystem $B$ is unchanged.
In fact, one can show that even the initial $\gamma N$ Bell pairs can be reconstructed by local operations in $A_1$ with an exponentially good precision~\cite{schumacher2002approximate}.

The inequality in Eq.~\eqref{eqn:criterion} is enough to prove the stability of volume-law scaling of the entanglement entropy in the presence of extensive number of measurements. However, it is not tight. 
This is because in deriving the inequality we assumed that qubits in $A_2$ are discarded (i.e., qubit loss errors), whereas in our situation they are projectively measured, leaving their measurement outcomes as \emph{accessible classical information}.
In~\cite{SOM}, we develop an improved decoupling equality, in which the measurement outcomes from $A_2$ are treated as accessible information. This leads to a tight bound
\begin{align}\label{eqn:tight_bound}
    \gamma  < 1-p \end{align}
in the limit $N \to \infty$.
This inequality can be saturated by typical Haar random unitaries.
An intuitive way to understand this new result is to realize that the measurement processes of $pN$ qubits involve entangling those qubits with an equal number of auxiliary qubits representing the environment (or classical measurement devices). These additional degrees of freedom effectively add to the right-hand side of Eq.~\eqref{eqn:criterion}, i.e., $\gamma+p<(1+\gamma+p)/2$, leading to the tight bound.

\begin{figure*}[t!]
\centering
\includegraphics[width=0.98\textwidth]{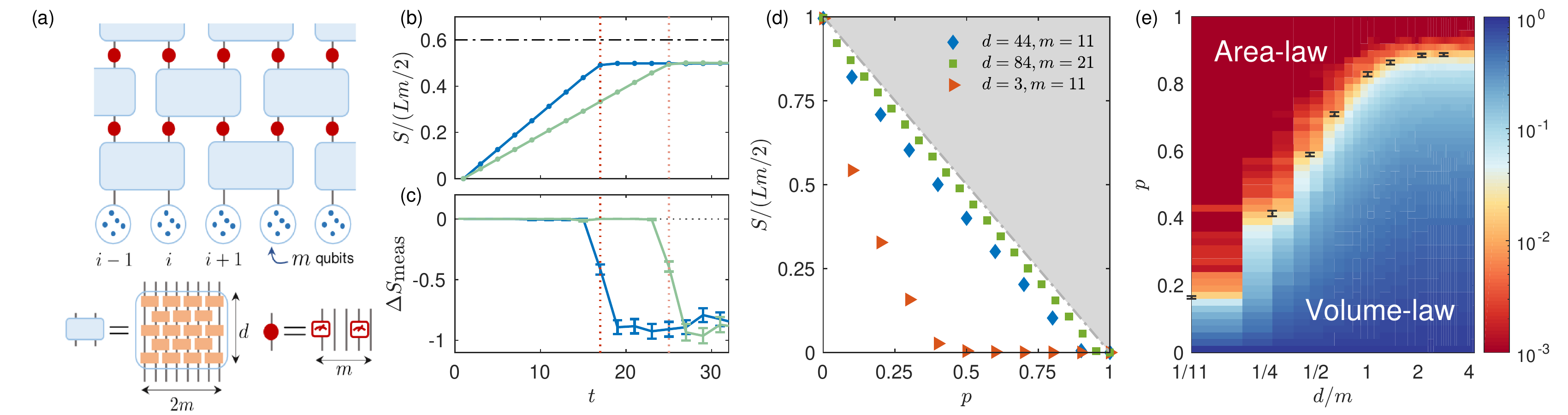}
\caption{
(a) A model with tunable degrees of information scrambling $d$ and  measurements $p$.
An array of $m$-qubit blocks undergoes layers of unitary gates (light blue) and random projective measurements (red).
Each unitary acting on neighboring blocks comprises independently random 2-qubit gates (orange).
Each measurement projects a randomly chosen $p$ fraction of qubits in each block.
(b,c) Entanglement dynamics with $m$\,$=$\,$11$, $d$\,$=$\,$44$, and $p$\,$=$\,$0.4$ for two different system sizes $L$\,$=$\,$32$ (blue) and $48$ (light green). 
 (b) The growth of entanglement density as a function of time $t$. The dash-dotted line indicates the upper bound $1-p$.
 (c) Change in the entanglement entropy before and after projective measurements at each time step.
 (d) Steady-state entanglement entropy per qubit as a function of $p$ for $(d, m)$\,$=$\,$(44, 11)$, $(84, 21)$, and $(3, 11)$.
 (e) Phase diagram for $m = 11$.
The color-coded background displays the half-chain entanglement entropies in steady states, normalized by the number of qubits $Lm/2$\,$=$\,$176$.
Black markers indicate the phase transition points extracted from finite size scaling analysis up to $L = 64$.
The numerical results in (b)-(e) are averaged over $240$ different realizations of random circuits and measurements.
} 
\label{fig:wide}
\end{figure*}

So far, we have considered an ideal situation where the Bell pairs are hidden over the entire Hilbert space via a nonlocal unitary. However, we emphasize that such information scrambling is a generic property of quantum dynamics even in local systems~\cite{Brandao2016,Haah2017entanglement,von2018operator,nahum2018operator,jonay2018coarse}.
In such cases, we expect that the amount of entanglement reduction is governed by the competition between the rate of effective information scrambling and that of projective measurements.

\emph{Model and phase diagram}.---Having understood the mechanism to protect the entanglement against measurements through scrambling, we turn to study a local 1D model in which the rates of effective information scrambling and measurements can be tuned independently.

Our model  consists of a chain of $L$ blocks, each containing a fixed number, $m$\,$\gg$\,$1$, of qubits, as illustrated in Fig.~\ref{fig:wide}(a).
In each time step $t$, the system is evolved by a network of random unitaries $U_d(i,t)$ acting on pairs of neighboring blocks at $i$ and $i+1$, supplemented by projective measurements.
Crucially, the unitaries $U_d(i,t)$  are constructed from an internal network consisting of $d$ layers of independent random 2-qubit gates (drawn from any unitary 2-design).
Thus, the parameter $d$ controls the degree of information scrambling within a single $U_d(i,t)$, which becomes maximally scrambling in the limit $d/m$\,$\gg$\,$1$. In this limit the distribution of  $U_d(i,t)$ approaches a unitary 2-design over U$(2^{2m})$~\cite{Brandao2016}.
After applications of the $U_d(i,t)$ on pairs of blocks, a fraction $p$ of the qubits in each block is randomly chosen to be measured in the computational basis~\cite{f_noninteger_p}.
We note that the special case of our model, $d$\,$=$\,$1$, is closely related to the previously studied ones~\cite{LiFisher2018ZenoEffect,SkinnerNahum2018measure,chan2019unitary}.

Before obtaining a quantitative phase diagram from numerical simulations [Fig.~\ref{fig:wide}(e)], one can  already predict the stability of volume-law phase in the limit $m$\,$\gg$\,$1$ and $d/m$\,$\gg$\,$1$. 
Consider the unitary evolution $U_d (i,t)$ for a pair of blocks.
If we identify the pair of blocks as subsystem $A$ and the rest of the system as $B$, we can use the decoupling inequality as discussed above.
As long as the average entropy per qubit $\gamma$ satisfies the criteria in Eq.~\eqref{eqn:tight_bound}, the measured qubits contain almost no information about the rest of the system (up to corrections exponentially small in $m$).
Here, the entanglement reduction is suppressed by information scrambling within the blocks.
Over multiple time steps, quantum information becomes scrambled over a larger region, further protecting the entanglement from measurements.
Thus we expect a stable volume-law phase in this regime.

We can also make a definitive statement about the other extreme of small $d$ and high measurement rate.
For example, consider $d$\,$=$\,$1$, $m$\,$\gg$\,$1$ and $p$\,$=$\,$1-1/m$. In this case there is no room for scrambling; thus, the probability that a single qubit becomes entangled to other qubits at distance $x$ away (in units of qubit blocks) is exponentially suppressed as $\sim$\,$(1/m)^{mx}$ because the information encoded in the qubit needs to propagate without being projected at least $\sim$\,$mx$ time steps. 
This implies  area-law entanglement~\cite{cao2019entanglement}. 
Therefore, we expect a phase transition between the two extreme cases.

We now complement the theoretical arguments with, 
a numerical simulation of the half-chain entanglement entropy $S(t)$ starting from the initial state  $\ket{\Psi_0}$\,$=$\,$\ket{0}^{\otimes mL}$.
We construct the unitary operators $U_d(i,t)$ from random 2-qubit Clifford gates drawn from a uniform distribution instead of taking Haar random unitaries.
Such $U_d(i,t)$ still approaches a unitary 2-design as $d$ increases~\cite{SOM};
hence, this simplification does not affect our predictions, while allowing scalable numerical simulation~\cite{gottesman1998heisenberg,aaronson2004improved,hamma2005ground,hamma2005bipartite}.
Furthermore, the wave function evolved under Clifford gates always exhibits a flat entanglement spectrum with respect to any bipartition.
Thus, different measures of entanglement entropy, e.g. von Neumann versus R\'enyi entropies, yield the same value.
In the following, we focus on $m$\,$=$\,$11$.

We first consider a strongly scrambling regime, $d/(2m)$\,$\gtrsim$\,$1$, where the unitary network within a single block effectively acts as a random $2^{2m}$\,$\times$\,$2^{2m}$ unitary~\cite{Brandao2016,SOM}. We test if the scrambling property of individual blocks leads to the robust volume-law entanglement of the entire system.
Figures~\ref{fig:wide}(b) and~\ref{fig:wide}(c) show the detailed dynamics of the entanglement entropy for two different system sizes. 
Clearly, the entropy rescaled by the subsystem size exhibits a strict linear growth until it saturates to a constant value.
The convergence of these values confirms the volume-law scaling of the entropy~\cite{SOM}. 
Moreover, we can directly compute how much the entanglement entropy changes, $\Delta S_\textrm{meas}(t)$, following projective measurements in each time step.
Figure~\ref{fig:wide}(c) confirms our prediction that in the strongly scrambling regime $d/(2m)\gtrsim 1$ and $m\gg 1$, the entanglement is unchanged by measurements until it reaches a saturation value set by the maximal entanglement that can be protected by the scrambling dynamics~\cite{footnoteOddStep}. 
Once saturated, the entanglement added by the unitary gates pushes the entropy above the threshold of the decoupling theorem and it is reduced back to the saturation value by the subsequent measurements. Thus, upon reaching the saturation value we see a jump of $\Delta S_{\textrm{meas}}$ from near zero to a negative value.
We further note that the saturation value approaches its maximum, $1-p$, as $m$ is increased in this regime [Fig.~\ref{fig:wide}(d)].
This is natural since our tight bound in Eq.~\eqref{eqn:tight_bound} becomes exact when $m\to \infty$, and it predicts that each qubit on average contributes $\gamma = 1-p$ to the global entanglement.
We note that this analysis does not hold in the weakly scrambling regime $d/(2m) \lesssim 1$, e.g., $d$\,$=$\,$3$, $m$\,$=$\,$11$ [Fig.~\ref{fig:wide}(d)].

We now turn to the phase transition that occurs when $d$ is decreased or $p$ is increased.
From numerical simulations, we compute the half-chain entanglement per qubit and tripartite mutual information in the steady state for various $L$ and $p$ for a fixed $d$~\cite{SOM}.
We perform a finite size scaling analysis in order to extract the critical measurement fraction $p_c$ as well as the correlation length critical exponent $\nu$~\cite{SkinnerNahum2018measure,gullans2019dynamical}.
By repeating the analysis for various values of $d$, we obtain a two-dimensional phase diagram shown in Fig.~\ref{fig:wide}(e).
We find that the fitted critical exponent $\nu$ has a universal value around $1.2$ independent of $d$ and $m$, suggesting the universality of the transition ~\cite{SOM}.
When $d$\,$=$\,$1$ the extracted critical value $p_c$\,$\approx$\,$0.16$ is consistent with previously reported results~\cite{LiFisher2018ZenoEffect,li2019measurement}.
More importantly, we find that the volume-law entangled phase extends to a higher measurement fraction, as $d$ increases to $\sim$\,$m$, and then saturates for $d/(2m)$\,$\gtrsim$\,$1$.

\emph{Discussion.---}
The existence of the stable volume-law phase has a direct interpretation in terms of QEC for quantum communications~\cite{nielsen2002quantum}, where the primary goal is to devise an encoding scheme to transfer the maximum amount of quantum information over a noisy or lossy channel. 
The maximum amount of coherent quantum information that can be transmitted through such a  channel is called the quantum channel capacity $\mathcal{Q}$~\cite{PreskillNotes,wilde2013quantum}.
Previously, one of the most important applications of the decoupling theorem had been to show that, by using a random unitary encoding, it is possible to asymptotically transfer $1-2p$ logical qubits per physical qubit over a lossy channel in which a fraction $p$ of the physical qubits is lost~\cite{PreskillNotes}.
In our settings, the projective measurements are distinguished from qubit loss errors since their measurement outcomes are available as classical information. This allows achieving a higher quantum channel capacity $1-p$ that we prove using a new decoupling inequality~\cite{SOM}.

The connection between the quantum channel capacity and the volume-law phase can be made more precise in two different settings.
In the specific setting of our 1D model, we considered the capacity within a pair of neighboring qubit blocks.
Here, the quantum information we wish to protect is quantified by the entanglement entropy between the qubit blocks and the rest of the system.
The random unitary circuit is equivalent to repeated encoding of the information without explicit decoding.
Since this encoding scheme can protect $\sim (1-p)2m$ logical qubits in each pair of $m$-qubit blocks~\cite{SOM}, we expect that our system should exhibit a stable volume-law scaling of entanglement supported by those logical qubits.

In a more general setting, we consider the entire system dynamics as a quantum channel and investigate its quantum channel capacity $\mathcal{Q}$.
To this end, we take the input state to be entangled with an auxiliary reference such that its reduced density matrix is $\rho_\textrm{in}$.
Then, we ask how much entanglement with the reference can be recovered from the combination of the output system density matrix $\rho_\textrm{out}$ and a set of classical measurement outcomes after a long time evolution.
This can be quantified by the \emph{coherent information} $\mathcal{I}_c(\mathcal{N},\rho_\textrm{in})$~\cite{devetak2005capacity}.  
For a quantum channel $\mathcal{N}$ consisting of unitary evolution interspersed with measurements (in any positive-operator valued measures), we show that~\cite{SOM}
\begin{align}
\label{eqn:channel_capacity}
    \mathcal{Q} 
    = \max_{\rho_\textrm{in}} \mathcal{I}_c\left(\mathcal{N}, \rho_\textrm{in}\right) 
    =  \max_{\rho_\textrm{in}} \langle S(\rho_{\text{out}}) \rangle
\end{align}
where $\langle S(\rho_{\text{out}}) \rangle$ is the von~Neumann entropy of $\rho_\textrm{out}$, averaged over all possible measurement outcomes during the time evolution.
We note that the first equality in Eq.~\eqref{eqn:channel_capacity} is not generically valid, but it holds for a class of so-called \emph{degradable} quantum channels that include our cases~\cite{devetak2005capacity,PreskillNotes,wilde2013quantum}.
Using the second equality, we establish an exact connection between quantum channel capacity and the average entropy of a system undergoing any unitary evolution interspersed by measurements.
A similar connection was first suggested in a recent paper by Gullans and Huse~\cite{gullans2019dynamical}.

In Ref.~\cite{gullans2019dynamical},  $\langle S(\rho_\textrm{out})\rangle$ for the maximally mixed initial state was shown to undergo a phase transition, which coincides with the entanglement phase transition for pure initial states.
Specifically, 
in the volume-law phase, $\langle S(\rho_{\text{out}}) \rangle $ remains extensive at late time, while in the area-law phase, it rapidly approaches a value of order one.
This transition was dubbed the purification phase transition~\cite{gullans2019dynamical}.
The equivalence between the purification and entanglement phase transitions has later been established analytically for local Haar random unitary circuits with measurements~\cite{bao2019theory}.
Therefore, Eq.~\eqref{eqn:channel_capacity} also builds a quantitative connection between the quantum channel capacity and the entanglement phase transition.
We note that, in random circuit models studied in Refs.~\cite{SkinnerNahum2018measure,LiFisher2018ZenoEffect,gullans2019dynamical,bao2019theory,jian2019measurement}, the channel $\mathcal{N}$ itself is random, whose average $\mathcal{I}_c$ is maximized for the maximally mixed input, furthering the connection~\cite{SOM}.

\begin{acknowledgments}
We note that, after we conjectured the relation between the quantum channel capacity and entanglement phase transition in the earlier preprint version of this paper, Ref.~\cite{gullans2019dynamical} provided the first quantitative evidence, proving that the single-shot quantum channel capacity is upper bounded by the average entropy of the output and showed that the stabilizer circuit saturates the bound. In this new version, we further prove a generic equality between output entropy and quantum channel capacity by considering the classical information in measurement outcomes as part of the channel.

We thank M. Gullans, D. Huse, X. Chen, Y. Li, I. Kim, A. Nahum, and Q. Zhuang for useful discussions.
S. C. acknowledges support from the Miller Institute for Basic Research in Science. Y. B. acknowledges support from the GSI program at UC Berkeley. X.-L. Q. and E. A. are supported in part by the Department of Energy Project No. DE-SC0019380. E. A. acknowledges support from the ERC synergy grant UQUAM and from the Gyorgy Chair in Physics at UC Berkeley.
\end{acknowledgments}
\bibliography{refs}

%merlin.mbs apsrev4-1.bst 2010-07-25 4.21a (PWD, AO, DPC) hacked
%Control: key (0)
%Control: author (8) initials jnrlst
%Control: editor formatted (1) identically to author
%Control: production of article title (-1) disabled
%Control: page (0) single
%Control: year (1) truncated
%Control: production of eprint (0) enabled
\begin{thebibliography}{46}%
\makeatletter
\providecommand \@ifxundefined [1]{%
 \@ifx{#1\undefined}
}%
\providecommand \@ifnum [1]{%
 \ifnum #1\expandafter \@firstoftwo
 \else \expandafter \@secondoftwo
 \fi
}%
\providecommand \@ifx [1]{%
 \ifx #1\expandafter \@firstoftwo
 \else \expandafter \@secondoftwo
 \fi
}%
\providecommand \natexlab [1]{#1}%
\providecommand \enquote  [1]{``#1''}%
\providecommand \bibnamefont  [1]{#1}%
\providecommand \bibfnamefont [1]{#1}%
\providecommand \citenamefont [1]{#1}%
\providecommand \href@noop [0]{\@secondoftwo}%
\providecommand \href [0]{\begingroup \@sanitize@url \@href}%
\providecommand \@href[1]{\@@startlink{#1}\@@href}%
\providecommand \@@href[1]{\endgroup#1\@@endlink}%
\providecommand \@sanitize@url [0]{\catcode `\\12\catcode `\$12\catcode
  `\&12\catcode `\#12\catcode `\^12\catcode `\_12\catcode `\%12\relax}%
\providecommand \@@startlink[1]{}%
\providecommand \@@endlink[0]{}%
\providecommand \url  [0]{\begingroup\@sanitize@url \@url }%
\providecommand \@url [1]{\endgroup\@href {#1}{\urlprefix }}%
\providecommand \urlprefix  [0]{URL }%
\providecommand \Eprint [0]{\href }%
\providecommand \doibase [0]{http://dx.doi.org/}%
\providecommand \selectlanguage [0]{\@gobble}%
\providecommand \bibinfo  [0]{\@secondoftwo}%
\providecommand \bibfield  [0]{\@secondoftwo}%
\providecommand \translation [1]{[#1]}%
\providecommand \BibitemOpen [0]{}%
\providecommand \bibitemStop [0]{}%
\providecommand \bibitemNoStop [0]{.\EOS\space}%
\providecommand \EOS [0]{\spacefactor3000\relax}%
\providecommand \BibitemShut  [1]{\csname bibitem#1\endcsname}%
\let\auto@bib@innerbib\@empty
%</preamble>
\bibitem [{\citenamefont {Deutsch}(1991)}]{DeutschETH}%
  \BibitemOpen
  \bibfield  {author} {\bibinfo {author} {\bibfnamefont {J.~M.}\ \bibnamefont
  {Deutsch}},\ }\href {\doibase 10.1103/PhysRevA.43.2046} {\bibfield  {journal}
  {\bibinfo  {journal} {Phys. Rev. A}\ }\textbf {\bibinfo {volume} {43}},\
  \bibinfo {pages} {2046} (\bibinfo {year} {1991})}\BibitemShut {NoStop}%
\bibitem [{\citenamefont {Srednicki}(1994)}]{SrednickiETH}%
  \BibitemOpen
  \bibfield  {author} {\bibinfo {author} {\bibfnamefont {M.}~\bibnamefont
  {Srednicki}},\ }\href {\doibase 10.1103/PhysRevE.50.888} {\bibfield
  {journal} {\bibinfo  {journal} {Phys. Rev. E}\ }\textbf {\bibinfo {volume}
  {50}},\ \bibinfo {pages} {888} (\bibinfo {year} {1994})}\BibitemShut
  {NoStop}%
\bibitem [{\citenamefont {Rigol}\ \emph {et~al.}(2008)\citenamefont {Rigol},
  \citenamefont {Dunjko},\ and\ \citenamefont
  {Olshanii}}]{rigol2008thermalization}%
  \BibitemOpen
  \bibfield  {author} {\bibinfo {author} {\bibfnamefont {M.}~\bibnamefont
  {Rigol}}, \bibinfo {author} {\bibfnamefont {V.}~\bibnamefont {Dunjko}}, \
  and\ \bibinfo {author} {\bibfnamefont {M.}~\bibnamefont {Olshanii}},\
  }\href@noop {} {\bibfield  {journal} {\bibinfo  {journal} {Nature}\ }\textbf
  {\bibinfo {volume} {452}},\ \bibinfo {pages} {854} (\bibinfo {year}
  {2008})}\BibitemShut {NoStop}%
\bibitem [{\citenamefont {Hayden}\ and\ \citenamefont
  {Preskill}(2007)}]{hayden2007black}%
  \BibitemOpen
  \bibfield  {author} {\bibinfo {author} {\bibfnamefont {P.}~\bibnamefont
  {Hayden}}\ and\ \bibinfo {author} {\bibfnamefont {J.}~\bibnamefont
  {Preskill}},\ }\href@noop {} {\bibfield  {journal} {\bibinfo  {journal}
  {Journal of High Energy Physics}\ }\textbf {\bibinfo {volume} {2007}},\
  \bibinfo {pages} {120} (\bibinfo {year} {2007})}\BibitemShut {NoStop}%
\bibitem [{\citenamefont {Sekino}\ and\ \citenamefont
  {Susskind}(2008)}]{sekino2008fast}%
  \BibitemOpen
  \bibfield  {author} {\bibinfo {author} {\bibfnamefont {Y.}~\bibnamefont
  {Sekino}}\ and\ \bibinfo {author} {\bibfnamefont {L.}~\bibnamefont
  {Susskind}},\ }\href@noop {} {\bibfield  {journal} {\bibinfo  {journal}
  {Journal of High Energy Physics}\ }\textbf {\bibinfo {volume} {2008}},\
  \bibinfo {pages} {065} (\bibinfo {year} {2008})}\BibitemShut {NoStop}%
\bibitem [{\citenamefont {Shenker}\ and\ \citenamefont
  {Stanford}(2014)}]{shenker2014black}%
  \BibitemOpen
  \bibfield  {author} {\bibinfo {author} {\bibfnamefont {S.~H.}\ \bibnamefont
  {Shenker}}\ and\ \bibinfo {author} {\bibfnamefont {D.}~\bibnamefont
  {Stanford}},\ }\href@noop {} {\bibfield  {journal} {\bibinfo  {journal}
  {Journal of High Energy Physics}\ }\textbf {\bibinfo {volume} {2014}},\
  \bibinfo {pages} {67} (\bibinfo {year} {2014})}\BibitemShut {NoStop}%
\bibitem [{\citenamefont {Hosur}\ \emph {et~al.}(2016)\citenamefont {Hosur},
  \citenamefont {Qi}, \citenamefont {Roberts},\ and\ \citenamefont
  {Yoshida}}]{hosur2016chaos}%
  \BibitemOpen
  \bibfield  {author} {\bibinfo {author} {\bibfnamefont {P.}~\bibnamefont
  {Hosur}}, \bibinfo {author} {\bibfnamefont {X.-L.}\ \bibnamefont {Qi}},
  \bibinfo {author} {\bibfnamefont {D.~A.}\ \bibnamefont {Roberts}}, \ and\
  \bibinfo {author} {\bibfnamefont {B.}~\bibnamefont {Yoshida}},\ }\href@noop
  {} {\bibfield  {journal} {\bibinfo  {journal} {Journal of High Energy
  Physics}\ }\textbf {\bibinfo {volume} {2016}},\ \bibinfo {pages} {4}
  (\bibinfo {year} {2016})}\BibitemShut {NoStop}%
\bibitem [{\citenamefont {Nahum}\ \emph {et~al.}(2017)\citenamefont {Nahum},
  \citenamefont {Ruhman}, \citenamefont {Vijay},\ and\ \citenamefont
  {Haah}}]{Haah2017entanglement}%
  \BibitemOpen
  \bibfield  {author} {\bibinfo {author} {\bibfnamefont {A.}~\bibnamefont
  {Nahum}}, \bibinfo {author} {\bibfnamefont {J.}~\bibnamefont {Ruhman}},
  \bibinfo {author} {\bibfnamefont {S.}~\bibnamefont {Vijay}}, \ and\ \bibinfo
  {author} {\bibfnamefont {J.}~\bibnamefont {Haah}},\ }\href {\doibase
  10.1103/PhysRevX.7.031016} {\bibfield  {journal} {\bibinfo  {journal} {Phys.
  Rev. X}\ }\textbf {\bibinfo {volume} {7}},\ \bibinfo {pages} {031016}
  (\bibinfo {year} {2017})}\BibitemShut {NoStop}%
\bibitem [{\citenamefont {von Keyserlingk}\ \emph {et~al.}(2018)\citenamefont
  {von Keyserlingk}, \citenamefont {Rakovszky}, \citenamefont {Pollmann},\ and\
  \citenamefont {Sondhi}}]{von2018operator}%
  \BibitemOpen
  \bibfield  {author} {\bibinfo {author} {\bibfnamefont {C.~W.}\ \bibnamefont
  {von Keyserlingk}}, \bibinfo {author} {\bibfnamefont {T.}~\bibnamefont
  {Rakovszky}}, \bibinfo {author} {\bibfnamefont {F.}~\bibnamefont {Pollmann}},
  \ and\ \bibinfo {author} {\bibfnamefont {S.~L.}\ \bibnamefont {Sondhi}},\
  }\href@noop {} {\bibfield  {journal} {\bibinfo  {journal} {Physical Review
  X}\ }\textbf {\bibinfo {volume} {8}},\ \bibinfo {pages} {021013} (\bibinfo
  {year} {2018})}\BibitemShut {NoStop}%
\bibitem [{\citenamefont {Nahum}\ \emph {et~al.}(2018)\citenamefont {Nahum},
  \citenamefont {Vijay},\ and\ \citenamefont {Haah}}]{nahum2018operator}%
  \BibitemOpen
  \bibfield  {author} {\bibinfo {author} {\bibfnamefont {A.}~\bibnamefont
  {Nahum}}, \bibinfo {author} {\bibfnamefont {S.}~\bibnamefont {Vijay}}, \ and\
  \bibinfo {author} {\bibfnamefont {J.}~\bibnamefont {Haah}},\ }\href@noop {}
  {\bibfield  {journal} {\bibinfo  {journal} {Physical Review X}\ }\textbf
  {\bibinfo {volume} {8}},\ \bibinfo {pages} {021014} (\bibinfo {year}
  {2018})}\BibitemShut {NoStop}%
\bibitem [{\citenamefont {Jonay}\ \emph {et~al.}(2018)\citenamefont {Jonay},
  \citenamefont {Huse},\ and\ \citenamefont {Nahum}}]{jonay2018coarse}%
  \BibitemOpen
  \bibfield  {author} {\bibinfo {author} {\bibfnamefont {C.}~\bibnamefont
  {Jonay}}, \bibinfo {author} {\bibfnamefont {D.~A.}\ \bibnamefont {Huse}}, \
  and\ \bibinfo {author} {\bibfnamefont {A.}~\bibnamefont {Nahum}},\
  }\href@noop {} {\bibfield  {journal} {\bibinfo  {journal} {arXiv preprint
  arXiv:1803.00089}\ } (\bibinfo {year} {2018})}\BibitemShut {NoStop}%
\bibitem [{\citenamefont {Aharonov}(2000)}]{aharonov2000quantum}%
  \BibitemOpen
  \bibfield  {author} {\bibinfo {author} {\bibfnamefont {D.}~\bibnamefont
  {Aharonov}},\ }\href@noop {} {\bibfield  {journal} {\bibinfo  {journal}
  {Physical Review A}\ }\textbf {\bibinfo {volume} {62}},\ \bibinfo {pages}
  {062311} (\bibinfo {year} {2000})}\BibitemShut {NoStop}%
\bibitem [{\citenamefont {Vidal}(2004)}]{VidalMPS2004}%
  \BibitemOpen
  \bibfield  {author} {\bibinfo {author} {\bibfnamefont {G.}~\bibnamefont
  {Vidal}},\ }\href {\doibase 10.1103/PhysRevLett.93.040502} {\bibfield
  {journal} {\bibinfo  {journal} {Phys. Rev. Lett.}\ }\textbf {\bibinfo
  {volume} {93}},\ \bibinfo {pages} {040502} (\bibinfo {year}
  {2004})}\BibitemShut {NoStop}%
\bibitem [{\citenamefont {Cao}\ \emph {et~al.}(2019)\citenamefont {Cao},
  \citenamefont {Tilloy},\ and\ \citenamefont {De~Luca}}]{cao2019entanglement}%
  \BibitemOpen
  \bibfield  {author} {\bibinfo {author} {\bibfnamefont {X.}~\bibnamefont
  {Cao}}, \bibinfo {author} {\bibfnamefont {A.}~\bibnamefont {Tilloy}}, \ and\
  \bibinfo {author} {\bibfnamefont {A.}~\bibnamefont {De~Luca}},\ }\href@noop
  {} {\bibfield  {journal} {\bibinfo  {journal} {SciPost Physics}\ }\textbf
  {\bibinfo {volume} {7}},\ \bibinfo {pages} {024} (\bibinfo {year}
  {2019})}\BibitemShut {NoStop}%
\bibitem [{\citenamefont {Skinner}\ \emph {et~al.}(2019)\citenamefont
  {Skinner}, \citenamefont {Ruhman},\ and\ \citenamefont
  {Nahum}}]{SkinnerNahum2018measure}%
  \BibitemOpen
  \bibfield  {author} {\bibinfo {author} {\bibfnamefont {B.}~\bibnamefont
  {Skinner}}, \bibinfo {author} {\bibfnamefont {J.}~\bibnamefont {Ruhman}}, \
  and\ \bibinfo {author} {\bibfnamefont {A.}~\bibnamefont {Nahum}},\
  }\href@noop {} {\bibfield  {journal} {\bibinfo  {journal} {Physical Review
  X}\ }\textbf {\bibinfo {volume} {9}},\ \bibinfo {pages} {031009} (\bibinfo
  {year} {2019})}\BibitemShut {NoStop}%
\bibitem [{\citenamefont {Li}\ \emph {et~al.}(2018)\citenamefont {Li},
  \citenamefont {Chen},\ and\ \citenamefont {Fisher}}]{LiFisher2018ZenoEffect}%
  \BibitemOpen
  \bibfield  {author} {\bibinfo {author} {\bibfnamefont {Y.}~\bibnamefont
  {Li}}, \bibinfo {author} {\bibfnamefont {X.}~\bibnamefont {Chen}}, \ and\
  \bibinfo {author} {\bibfnamefont {M.~P.~A.}\ \bibnamefont {Fisher}},\ }\href
  {\doibase 10.1103/PhysRevB.98.205136} {\bibfield  {journal} {\bibinfo
  {journal} {Phys. Rev. B}\ }\textbf {\bibinfo {volume} {98}},\ \bibinfo
  {pages} {205136} (\bibinfo {year} {2018})}\BibitemShut {NoStop}%
\bibitem [{\citenamefont {Chan}\ \emph {et~al.}(2019)\citenamefont {Chan},
  \citenamefont {Nandkishore}, \citenamefont {Pretko},\ and\ \citenamefont
  {Smith}}]{chan2019unitary}%
  \BibitemOpen
  \bibfield  {author} {\bibinfo {author} {\bibfnamefont {A.}~\bibnamefont
  {Chan}}, \bibinfo {author} {\bibfnamefont {R.~M.}\ \bibnamefont
  {Nandkishore}}, \bibinfo {author} {\bibfnamefont {M.}~\bibnamefont {Pretko}},
  \ and\ \bibinfo {author} {\bibfnamefont {G.}~\bibnamefont {Smith}},\
  }\href@noop {} {\bibfield  {journal} {\bibinfo  {journal} {Physical Review
  B}\ }\textbf {\bibinfo {volume} {99}},\ \bibinfo {pages} {224307} (\bibinfo
  {year} {2019})}\BibitemShut {NoStop}%
\bibitem [{\citenamefont {Nielsen}\ and\ \citenamefont
  {Chuang}(2000)}]{nielsen2002quantum}%
  \BibitemOpen
  \bibfield  {author} {\bibinfo {author} {\bibfnamefont {M.~A.}\ \bibnamefont
  {Nielsen}}\ and\ \bibinfo {author} {\bibfnamefont {I.~L.}\ \bibnamefont
  {Chuang}},\ }\href@noop {} {\emph {\bibinfo {title} {Quantum Computation and
  Quantum Information}}}\ (\bibinfo  {publisher} {Cambridge University Press},\
  \bibinfo {year} {2000})\BibitemShut {NoStop}%
\bibitem [{\citenamefont {Preskill}(2018)}]{PreskillNotes}%
  \BibitemOpen
  \bibfield  {author} {\bibinfo {author} {\bibfnamefont {J.}~\bibnamefont
  {Preskill}},\ }\href@noop {} {\enquote {\bibinfo {title} {Lecture notes for
  physics 219: Quantum computation},}\ } (\bibinfo {year} {2018})\BibitemShut
  {NoStop}%
\bibitem [{\citenamefont {Schumacher}\ and\ \citenamefont
  {Westmoreland}(2002)}]{schumacher2002approximate}%
  \BibitemOpen
  \bibfield  {author} {\bibinfo {author} {\bibfnamefont {B.}~\bibnamefont
  {Schumacher}}\ and\ \bibinfo {author} {\bibfnamefont {M.~D.}\ \bibnamefont
  {Westmoreland}},\ }\href@noop {} {\bibfield  {journal} {\bibinfo  {journal}
  {Quantum Information Processing}\ }\textbf {\bibinfo {volume} {1}},\ \bibinfo
  {pages} {5} (\bibinfo {year} {2002})}\BibitemShut {NoStop}%
\bibitem [{\citenamefont {Devetak}(2005)}]{devetak2005private}%
  \BibitemOpen
  \bibfield  {author} {\bibinfo {author} {\bibfnamefont {I.}~\bibnamefont
  {Devetak}},\ }\href@noop {} {\bibfield  {journal} {\bibinfo  {journal} {IEEE
  Transactions on Information Theory}\ }\textbf {\bibinfo {volume} {51}},\
  \bibinfo {pages} {44} (\bibinfo {year} {2005})}\BibitemShut {NoStop}%
\bibitem [{\citenamefont {Horodecki}\ \emph {et~al.}(2007)\citenamefont
  {Horodecki}, \citenamefont {Oppenheim},\ and\ \citenamefont
  {Winter}}]{horodecki2007quantum}%
  \BibitemOpen
  \bibfield  {author} {\bibinfo {author} {\bibfnamefont {M.}~\bibnamefont
  {Horodecki}}, \bibinfo {author} {\bibfnamefont {J.}~\bibnamefont
  {Oppenheim}}, \ and\ \bibinfo {author} {\bibfnamefont {A.}~\bibnamefont
  {Winter}},\ }\href@noop {} {\bibfield  {journal} {\bibinfo  {journal}
  {Communications in Mathematical Physics}\ }\textbf {\bibinfo {volume}
  {269}},\ \bibinfo {pages} {107} (\bibinfo {year} {2007})}\BibitemShut
  {NoStop}%
\bibitem [{\citenamefont {Abeyesinghe}\ \emph {et~al.}(2009)\citenamefont
  {Abeyesinghe}, \citenamefont {Devetak}, \citenamefont {Hayden},\ and\
  \citenamefont {Winter}}]{abeyesinghe2009mother}%
  \BibitemOpen
  \bibfield  {author} {\bibinfo {author} {\bibfnamefont {A.}~\bibnamefont
  {Abeyesinghe}}, \bibinfo {author} {\bibfnamefont {I.}~\bibnamefont
  {Devetak}}, \bibinfo {author} {\bibfnamefont {P.}~\bibnamefont {Hayden}}, \
  and\ \bibinfo {author} {\bibfnamefont {A.}~\bibnamefont {Winter}},\
  }\href@noop {} {\bibfield  {journal} {\bibinfo  {journal} {Proceedings of the
  Royal Society A: Mathematical, Physical and Engineering Sciences}\ }\textbf
  {\bibinfo {volume} {465}},\ \bibinfo {pages} {2537} (\bibinfo {year}
  {2009})}\BibitemShut {NoStop}%
\bibitem [{SOM()}]{SOM}%
  \BibitemOpen
  \href@noop {} {}\bibinfo {note} {See Supplementary Online Material for the
  detailed information on numerical simulations, the finite-size scaling
  analysis, the relation between the quantum channel capacity and the
  entanglement phase transition, and the derivation of the improved decoupling
  inequality, which includes
  Refs.~\cite{divincenzo2002quantum,roberts2017chaos,bennett1996mixed,calderbank1997quantum,gottesman1996class,webb2015clifford,houdayer2004low,kawashima1993critical,collins2003moments}}\BibitemShut
  {NoStop}%
\bibitem [{\citenamefont {Brand{\~a}o}\ \emph {et~al.}(2016)\citenamefont
  {Brand{\~a}o}, \citenamefont {Harrow},\ and\ \citenamefont
  {Horodecki}}]{Brandao2016}%
  \BibitemOpen
  \bibfield  {author} {\bibinfo {author} {\bibfnamefont {F.~G. S.~L.}\
  \bibnamefont {Brand{\~a}o}}, \bibinfo {author} {\bibfnamefont {A.~W.}\
  \bibnamefont {Harrow}}, \ and\ \bibinfo {author} {\bibfnamefont
  {M.}~\bibnamefont {Horodecki}},\ }\href {\doibase 10.1007/s00220-016-2706-8}
  {\bibfield  {journal} {\bibinfo  {journal} {Communications in Mathematical
  Physics}\ }\textbf {\bibinfo {volume} {346}},\ \bibinfo {pages} {397}
  (\bibinfo {year} {2016})}\BibitemShut {NoStop}%
\bibitem [{f_n()}]{f_noninteger_p}%
  \BibitemOpen
  \href@noop {} {}\bibinfo {note} {For noninteger $pm$, the number of measured
  qubits is determined from a binomial distribution between $\lfloor pm
  \rfloor$ and $\lceil pm \rceil$ with mean $pm$.}\BibitemShut {Stop}%
\bibitem [{\citenamefont {Gottesman}(1998)}]{gottesman1998heisenberg}%
  \BibitemOpen
  \bibfield  {author} {\bibinfo {author} {\bibfnamefont {D.}~\bibnamefont
  {Gottesman}},\ }\href@noop {} {\bibfield  {journal} {\bibinfo  {journal}
  {arXiv preprint quant-ph/9807006}\ } (\bibinfo {year} {1998})}\BibitemShut
  {NoStop}%
\bibitem [{\citenamefont {Aaronson}\ and\ \citenamefont
  {Gottesman}(2004)}]{aaronson2004improved}%
  \BibitemOpen
  \bibfield  {author} {\bibinfo {author} {\bibfnamefont {S.}~\bibnamefont
  {Aaronson}}\ and\ \bibinfo {author} {\bibfnamefont {D.}~\bibnamefont
  {Gottesman}},\ }\href@noop {} {\bibfield  {journal} {\bibinfo  {journal}
  {Physical Review A}\ }\textbf {\bibinfo {volume} {70}},\ \bibinfo {pages}
  {052328} (\bibinfo {year} {2004})}\BibitemShut {NoStop}%
\bibitem [{\citenamefont {Hamma}\ \emph
  {et~al.}(2005{\natexlab{a}})\citenamefont {Hamma}, \citenamefont
  {Ionicioiu},\ and\ \citenamefont {Zanardi}}]{hamma2005ground}%
  \BibitemOpen
  \bibfield  {author} {\bibinfo {author} {\bibfnamefont {A.}~\bibnamefont
  {Hamma}}, \bibinfo {author} {\bibfnamefont {R.}~\bibnamefont {Ionicioiu}}, \
  and\ \bibinfo {author} {\bibfnamefont {P.}~\bibnamefont {Zanardi}},\
  }\href@noop {} {\bibfield  {journal} {\bibinfo  {journal} {Physics Letters
  A}\ }\textbf {\bibinfo {volume} {337}},\ \bibinfo {pages} {22} (\bibinfo
  {year} {2005}{\natexlab{a}})}\BibitemShut {NoStop}%
\bibitem [{\citenamefont {Hamma}\ \emph
  {et~al.}(2005{\natexlab{b}})\citenamefont {Hamma}, \citenamefont
  {Ionicioiu},\ and\ \citenamefont {Zanardi}}]{hamma2005bipartite}%
  \BibitemOpen
  \bibfield  {author} {\bibinfo {author} {\bibfnamefont {A.}~\bibnamefont
  {Hamma}}, \bibinfo {author} {\bibfnamefont {R.}~\bibnamefont {Ionicioiu}}, \
  and\ \bibinfo {author} {\bibfnamefont {P.}~\bibnamefont {Zanardi}},\
  }\href@noop {} {\bibfield  {journal} {\bibinfo  {journal} {Physical Review
  A}\ }\textbf {\bibinfo {volume} {71}},\ \bibinfo {pages} {022315} (\bibinfo
  {year} {2005}{\natexlab{b}})}\BibitemShut {NoStop}%
\bibitem [{foo()}]{footnoteOddStep}%
  \BibitemOpen
  \href@noop {} {}\bibinfo {note} {We note that, for this purpose, we have only
  considered odd time steps since in even time steps projective measurements
  destroy local entanglement within the qubit blocks that are generated by
  immediately preceding $U_d(i,t)$.}\BibitemShut {Stop}%
\bibitem [{\citenamefont {Gullans}\ and\ \citenamefont
  {Huse}(2019)}]{gullans2019dynamical}%
  \BibitemOpen
  \bibfield  {author} {\bibinfo {author} {\bibfnamefont {M.~J.}\ \bibnamefont
  {Gullans}}\ and\ \bibinfo {author} {\bibfnamefont {D.~A.}\ \bibnamefont
  {Huse}},\ }\href@noop {} {\bibfield  {journal} {\bibinfo  {journal} {arXiv
  preprint arXiv:1905.05195}\ } (\bibinfo {year} {2019})}\BibitemShut {NoStop}%
\bibitem [{\citenamefont {Li}\ \emph {et~al.}(2019)\citenamefont {Li},
  \citenamefont {Chen},\ and\ \citenamefont {Fisher}}]{li2019measurement}%
  \BibitemOpen
  \bibfield  {author} {\bibinfo {author} {\bibfnamefont {Y.}~\bibnamefont
  {Li}}, \bibinfo {author} {\bibfnamefont {X.}~\bibnamefont {Chen}}, \ and\
  \bibinfo {author} {\bibfnamefont {M.~P.~A.}\ \bibnamefont {Fisher}},\
  }\href@noop {} {\bibfield  {journal} {\bibinfo  {journal} {Physical Review
  B}\ }\textbf {\bibinfo {volume} {100}},\ \bibinfo {pages} {134306} (\bibinfo
  {year} {2019})}\BibitemShut {NoStop}%
\bibitem [{\citenamefont {Wilde}(2013)}]{wilde2013quantum}%
  \BibitemOpen
  \bibfield  {author} {\bibinfo {author} {\bibfnamefont {M.~M.}\ \bibnamefont
  {Wilde}},\ }\href@noop {} {\emph {\bibinfo {title} {Quantum information
  theory}}}\ (\bibinfo  {publisher} {Cambridge University Press},\ \bibinfo
  {year} {2013})\BibitemShut {NoStop}%
\bibitem [{\citenamefont {Devetak}\ and\ \citenamefont
  {Shor}(2005)}]{devetak2005capacity}%
  \BibitemOpen
  \bibfield  {author} {\bibinfo {author} {\bibfnamefont {I.}~\bibnamefont
  {Devetak}}\ and\ \bibinfo {author} {\bibfnamefont {P.~W.}\ \bibnamefont
  {Shor}},\ }\href@noop {} {\bibfield  {journal} {\bibinfo  {journal}
  {Communications in Mathematical Physics}\ }\textbf {\bibinfo {volume}
  {256}},\ \bibinfo {pages} {287} (\bibinfo {year} {2005})}\BibitemShut
  {NoStop}%
\bibitem [{\citenamefont {Bao}\ \emph {et~al.}(2020)\citenamefont {Bao},
  \citenamefont {Choi},\ and\ \citenamefont {Altman}}]{bao2019theory}%
  \BibitemOpen
  \bibfield  {author} {\bibinfo {author} {\bibfnamefont {Y.}~\bibnamefont
  {Bao}}, \bibinfo {author} {\bibfnamefont {S.}~\bibnamefont {Choi}}, \ and\
  \bibinfo {author} {\bibfnamefont {E.}~\bibnamefont {Altman}},\ }\href@noop {}
  {\bibfield  {journal} {\bibinfo  {journal} {Physical Review B}\ }\textbf
  {\bibinfo {volume} {101}},\ \bibinfo {pages} {104301} (\bibinfo {year}
  {2020})}\BibitemShut {NoStop}%
\bibitem [{\citenamefont {Jian}\ \emph {et~al.}(2020)\citenamefont {Jian},
  \citenamefont {You}, \citenamefont {Vasseur},\ and\ \citenamefont
  {Ludwig}}]{jian2019measurement}%
  \BibitemOpen
  \bibfield  {author} {\bibinfo {author} {\bibfnamefont {C.-M.}\ \bibnamefont
  {Jian}}, \bibinfo {author} {\bibfnamefont {Y.-Z.}\ \bibnamefont {You}},
  \bibinfo {author} {\bibfnamefont {R.}~\bibnamefont {Vasseur}}, \ and\
  \bibinfo {author} {\bibfnamefont {A.~W.~W.}\ \bibnamefont {Ludwig}},\
  }\href@noop {} {\bibfield  {journal} {\bibinfo  {journal} {Physical Review
  B}\ }\textbf {\bibinfo {volume} {101}},\ \bibinfo {pages} {104302} (\bibinfo
  {year} {2020})}\BibitemShut {NoStop}%
\bibitem [{\citenamefont {DiVincenzo}\ \emph {et~al.}(2002)\citenamefont
  {DiVincenzo}, \citenamefont {Leung},\ and\ \citenamefont
  {Terhal}}]{divincenzo2002quantum}%
  \BibitemOpen
  \bibfield  {author} {\bibinfo {author} {\bibfnamefont {D.~P.}\ \bibnamefont
  {DiVincenzo}}, \bibinfo {author} {\bibfnamefont {D.~W.}\ \bibnamefont
  {Leung}}, \ and\ \bibinfo {author} {\bibfnamefont {B.~M.}\ \bibnamefont
  {Terhal}},\ }\href@noop {} {\bibfield  {journal} {\bibinfo  {journal} {IEEE
  Transactions on Information Theory}\ }\textbf {\bibinfo {volume} {48}},\
  \bibinfo {pages} {580} (\bibinfo {year} {2002})}\BibitemShut {NoStop}%
\bibitem [{\citenamefont {Roberts}\ and\ \citenamefont
  {Yoshida}(2017)}]{roberts2017chaos}%
  \BibitemOpen
  \bibfield  {author} {\bibinfo {author} {\bibfnamefont {D.~A.}\ \bibnamefont
  {Roberts}}\ and\ \bibinfo {author} {\bibfnamefont {B.}~\bibnamefont
  {Yoshida}},\ }\href@noop {} {\bibfield  {journal} {\bibinfo  {journal}
  {Journal of High Energy Physics}\ }\textbf {\bibinfo {volume} {2017}},\
  \bibinfo {pages} {121} (\bibinfo {year} {2017})}\BibitemShut {NoStop}%
\bibitem [{\citenamefont {Bennett}\ \emph {et~al.}(1996)\citenamefont
  {Bennett}, \citenamefont {DiVincenzo}, \citenamefont {Smolin},\ and\
  \citenamefont {Wootters}}]{bennett1996mixed}%
  \BibitemOpen
  \bibfield  {author} {\bibinfo {author} {\bibfnamefont {C.~H.}\ \bibnamefont
  {Bennett}}, \bibinfo {author} {\bibfnamefont {D.~P.}\ \bibnamefont
  {DiVincenzo}}, \bibinfo {author} {\bibfnamefont {J.~A.}\ \bibnamefont
  {Smolin}}, \ and\ \bibinfo {author} {\bibfnamefont {W.~K.}\ \bibnamefont
  {Wootters}},\ }\href@noop {} {\bibfield  {journal} {\bibinfo  {journal}
  {Physical Review A}\ }\textbf {\bibinfo {volume} {54}},\ \bibinfo {pages}
  {3824} (\bibinfo {year} {1996})}\BibitemShut {NoStop}%
\bibitem [{\citenamefont {Calderbank}\ \emph {et~al.}(1997)\citenamefont
  {Calderbank}, \citenamefont {Rains}, \citenamefont {Shor},\ and\
  \citenamefont {Sloane}}]{calderbank1997quantum}%
  \BibitemOpen
  \bibfield  {author} {\bibinfo {author} {\bibfnamefont {A.~R.}\ \bibnamefont
  {Calderbank}}, \bibinfo {author} {\bibfnamefont {E.~M.}\ \bibnamefont
  {Rains}}, \bibinfo {author} {\bibfnamefont {P.~W.}\ \bibnamefont {Shor}}, \
  and\ \bibinfo {author} {\bibfnamefont {N.~J.~A.}\ \bibnamefont {Sloane}},\
  }\href@noop {} {\bibfield  {journal} {\bibinfo  {journal} {Physical Review
  Letters}\ }\textbf {\bibinfo {volume} {78}},\ \bibinfo {pages} {405}
  (\bibinfo {year} {1997})}\BibitemShut {NoStop}%
\bibitem [{\citenamefont {Gottesman}(1996)}]{gottesman1996class}%
  \BibitemOpen
  \bibfield  {author} {\bibinfo {author} {\bibfnamefont {D.}~\bibnamefont
  {Gottesman}},\ }\href@noop {} {\bibfield  {journal} {\bibinfo  {journal}
  {Physical Review A}\ }\textbf {\bibinfo {volume} {54}},\ \bibinfo {pages}
  {1862} (\bibinfo {year} {1996})}\BibitemShut {NoStop}%
\bibitem [{\citenamefont {Webb}(2015)}]{webb2015clifford}%
  \BibitemOpen
  \bibfield  {author} {\bibinfo {author} {\bibfnamefont {Z.}~\bibnamefont
  {Webb}},\ }\href@noop {} {\bibfield  {journal} {\bibinfo  {journal} {arXiv
  preprint arXiv:1510.02769}\ } (\bibinfo {year} {2015})}\BibitemShut {NoStop}%
\bibitem [{\citenamefont {Houdayer}\ and\ \citenamefont
  {Hartmann}(2004)}]{houdayer2004low}%
  \BibitemOpen
  \bibfield  {author} {\bibinfo {author} {\bibfnamefont {J.}~\bibnamefont
  {Houdayer}}\ and\ \bibinfo {author} {\bibfnamefont {A.~K.}\ \bibnamefont
  {Hartmann}},\ }\href@noop {} {\bibfield  {journal} {\bibinfo  {journal}
  {Physical Review B}\ }\textbf {\bibinfo {volume} {70}},\ \bibinfo {pages}
  {014418} (\bibinfo {year} {2004})}\BibitemShut {NoStop}%
\bibitem [{\citenamefont {Kawashima}\ and\ \citenamefont
  {Ito}(1993)}]{kawashima1993critical}%
  \BibitemOpen
  \bibfield  {author} {\bibinfo {author} {\bibfnamefont {N.}~\bibnamefont
  {Kawashima}}\ and\ \bibinfo {author} {\bibfnamefont {N.}~\bibnamefont
  {Ito}},\ }\href@noop {} {\bibfield  {journal} {\bibinfo  {journal} {Journal
  of the Physical Society of Japan}\ }\textbf {\bibinfo {volume} {62}},\
  \bibinfo {pages} {435} (\bibinfo {year} {1993})}\BibitemShut {NoStop}%
\bibitem [{\citenamefont {Collins}(2003)}]{collins2003moments}%
  \BibitemOpen
  \bibfield  {author} {\bibinfo {author} {\bibfnamefont {B.}~\bibnamefont
  {Collins}},\ }\href@noop {} {\bibfield  {journal} {\bibinfo  {journal}
  {International Mathematics Research Notices}\ }\textbf {\bibinfo {volume}
  {2003}},\ \bibinfo {pages} {953} (\bibinfo {year} {2003})}\BibitemShut
  {NoStop}%
\end{thebibliography}%


%merlin.mbs apsrev4-1.bst 2010-07-25 4.21a (PWD, AO, DPC) hacked
%Control: key (0)
%Control: author (8) initials jnrlst
%Control: editor formatted (1) identically to author
%Control: production of article title (-1) disabled
%Control: page (0) single
%Control: year (1) truncated
%Control: production of eprint (0) enabled
\begin{thebibliography}{20}%
\makeatletter
\providecommand \@ifxundefined [1]{%
 \@ifx{#1\undefined}
}%
\providecommand \@ifnum [1]{%
 \ifnum #1\expandafter \@firstoftwo
 \else \expandafter \@secondoftwo
 \fi
}%
\providecommand \@ifx [1]{%
 \ifx #1\expandafter \@firstoftwo
 \else \expandafter \@secondoftwo
 \fi
}%
\providecommand \natexlab [1]{#1}%
\providecommand \enquote  [1]{``#1''}%
\providecommand \bibnamefont  [1]{#1}%
\providecommand \bibfnamefont [1]{#1}%
\providecommand \citenamefont [1]{#1}%
\providecommand \href@noop [0]{\@secondoftwo}%
\providecommand \href [0]{\begingroup \@sanitize@url \@href}%
\providecommand \@href[1]{\@@startlink{#1}\@@href}%
\providecommand \@@href[1]{\endgroup#1\@@endlink}%
\providecommand \@sanitize@url [0]{\catcode `\\12\catcode `\$12\catcode
  `\&12\catcode `\#12\catcode `\^12\catcode `\_12\catcode `\%12\relax}%
\providecommand \@@startlink[1]{}%
\providecommand \@@endlink[0]{}%
\providecommand \url  [0]{\begingroup\@sanitize@url \@url }%
\providecommand \@url [1]{\endgroup\@href {#1}{\urlprefix }}%
\providecommand \urlprefix  [0]{URL }%
\providecommand \Eprint [0]{\href }%
\providecommand \doibase [0]{http://dx.doi.org/}%
\providecommand \selectlanguage [0]{\@gobble}%
\providecommand \bibinfo  [0]{\@secondoftwo}%
\providecommand \bibfield  [0]{\@secondoftwo}%
\providecommand \translation [1]{[#1]}%
\providecommand \BibitemOpen [0]{}%
\providecommand \bibitemStop [0]{}%
\providecommand \bibitemNoStop [0]{.\EOS\space}%
\providecommand \EOS [0]{\spacefactor3000\relax}%
\providecommand \BibitemShut  [1]{\csname bibitem#1\endcsname}%
\let\auto@bib@innerbib\@empty
%</preamble>
\bibitem [{\citenamefont {DiVincenzo}\ \emph {et~al.}(2002)\citenamefont
  {DiVincenzo}, \citenamefont {Leung},\ and\ \citenamefont
  {Terhal}}]{divincenzo2002quantum}%
  \BibitemOpen
  \bibfield  {author} {\bibinfo {author} {\bibfnamefont {D.~P.}\ \bibnamefont
  {DiVincenzo}}, \bibinfo {author} {\bibfnamefont {D.~W.}\ \bibnamefont
  {Leung}}, \ and\ \bibinfo {author} {\bibfnamefont {B.~M.}\ \bibnamefont
  {Terhal}},\ }\href@noop {} {\bibfield  {journal} {\bibinfo  {journal} {IEEE
  Transactions on Information Theory}\ }\textbf {\bibinfo {volume} {48}},\
  \bibinfo {pages} {580} (\bibinfo {year} {2002})}\BibitemShut {NoStop}%
\bibitem [{\citenamefont {Roberts}\ and\ \citenamefont
  {Yoshida}(2017)}]{roberts2017chaos}%
  \BibitemOpen
  \bibfield  {author} {\bibinfo {author} {\bibfnamefont {D.~A.}\ \bibnamefont
  {Roberts}}\ and\ \bibinfo {author} {\bibfnamefont {B.}~\bibnamefont
  {Yoshida}},\ }\href@noop {} {\bibfield  {journal} {\bibinfo  {journal}
  {Journal of High Energy Physics}\ }\textbf {\bibinfo {volume} {2017}},\
  \bibinfo {pages} {121} (\bibinfo {year} {2017})}\BibitemShut {NoStop}%
\bibitem [{\citenamefont {Bennett}\ \emph {et~al.}(1996)\citenamefont
  {Bennett}, \citenamefont {DiVincenzo}, \citenamefont {Smolin},\ and\
  \citenamefont {Wootters}}]{bennett1996mixed}%
  \BibitemOpen
  \bibfield  {author} {\bibinfo {author} {\bibfnamefont {C.~H.}\ \bibnamefont
  {Bennett}}, \bibinfo {author} {\bibfnamefont {D.~P.}\ \bibnamefont
  {DiVincenzo}}, \bibinfo {author} {\bibfnamefont {J.~A.}\ \bibnamefont
  {Smolin}}, \ and\ \bibinfo {author} {\bibfnamefont {W.~K.}\ \bibnamefont
  {Wootters}},\ }\href@noop {} {\bibfield  {journal} {\bibinfo  {journal}
  {Physical Review A}\ }\textbf {\bibinfo {volume} {54}},\ \bibinfo {pages}
  {3824} (\bibinfo {year} {1996})}\BibitemShut {NoStop}%
\bibitem [{\citenamefont {Calderbank}\ \emph {et~al.}(1997)\citenamefont
  {Calderbank}, \citenamefont {Rains}, \citenamefont {Shor},\ and\
  \citenamefont {Sloane}}]{calderbank1997quantum}%
  \BibitemOpen
  \bibfield  {author} {\bibinfo {author} {\bibfnamefont {A.~R.}\ \bibnamefont
  {Calderbank}}, \bibinfo {author} {\bibfnamefont {E.~M.}\ \bibnamefont
  {Rains}}, \bibinfo {author} {\bibfnamefont {P.~W.}\ \bibnamefont {Shor}}, \
  and\ \bibinfo {author} {\bibfnamefont {N.~J.}\ \bibnamefont {Sloane}},\
  }\href@noop {} {\bibfield  {journal} {\bibinfo  {journal} {Physical Review
  Letters}\ }\textbf {\bibinfo {volume} {78}},\ \bibinfo {pages} {405}
  (\bibinfo {year} {1997})}\BibitemShut {NoStop}%
\bibitem [{\citenamefont {Gottesman}(1996)}]{gottesman1996class}%
  \BibitemOpen
  \bibfield  {author} {\bibinfo {author} {\bibfnamefont {D.}~\bibnamefont
  {Gottesman}},\ }\href@noop {} {\bibfield  {journal} {\bibinfo  {journal}
  {Physical Review A}\ }\textbf {\bibinfo {volume} {54}},\ \bibinfo {pages}
  {1862} (\bibinfo {year} {1996})}\BibitemShut {NoStop}%
\bibitem [{\citenamefont {Gottesman}(1998)}]{gottesman1998heisenberg}%
  \BibitemOpen
  \bibfield  {author} {\bibinfo {author} {\bibfnamefont {D.}~\bibnamefont
  {Gottesman}},\ }\href@noop {} {\bibfield  {journal} {\bibinfo  {journal}
  {arXiv preprint quant-ph/9807006}\ } (\bibinfo {year} {1998})}\BibitemShut
  {NoStop}%
\bibitem [{\citenamefont {Aaronson}\ and\ \citenamefont
  {Gottesman}(2004)}]{aaronson2004improved}%
  \BibitemOpen
  \bibfield  {author} {\bibinfo {author} {\bibfnamefont {S.}~\bibnamefont
  {Aaronson}}\ and\ \bibinfo {author} {\bibfnamefont {D.}~\bibnamefont
  {Gottesman}},\ }\href@noop {} {\bibfield  {journal} {\bibinfo  {journal}
  {Physical Review A}\ }\textbf {\bibinfo {volume} {70}},\ \bibinfo {pages}
  {052328} (\bibinfo {year} {2004})}\BibitemShut {NoStop}%
\bibitem [{\citenamefont {Brandao}\ \emph {et~al.}(2016)\citenamefont
  {Brandao}, \citenamefont {Harrow},\ and\ \citenamefont
  {Horodecki}}]{brandao2016local}%
  \BibitemOpen
  \bibfield  {author} {\bibinfo {author} {\bibfnamefont {F.~G.}\ \bibnamefont
  {Brandao}}, \bibinfo {author} {\bibfnamefont {A.~W.}\ \bibnamefont {Harrow}},
  \ and\ \bibinfo {author} {\bibfnamefont {M.}~\bibnamefont {Horodecki}},\
  }\href@noop {} {\bibfield  {journal} {\bibinfo  {journal} {Communications in
  Mathematical Physics}\ }\textbf {\bibinfo {volume} {346}},\ \bibinfo {pages}
  {397} (\bibinfo {year} {2016})}\BibitemShut {NoStop}%
\bibitem [{\citenamefont {Webb}(2015)}]{webb2015clifford}%
  \BibitemOpen
  \bibfield  {author} {\bibinfo {author} {\bibfnamefont {Z.}~\bibnamefont
  {Webb}},\ }\href@noop {} {\bibfield  {journal} {\bibinfo  {journal} {arXiv
  preprint arXiv:1510.02769}\ } (\bibinfo {year} {2015})}\BibitemShut {NoStop}%
\bibitem [{\citenamefont {Li}\ \emph {et~al.}(2019)\citenamefont {Li},
  \citenamefont {Chen},\ and\ \citenamefont {Fisher}}]{li2019measurement}%
  \BibitemOpen
  \bibfield  {author} {\bibinfo {author} {\bibfnamefont {Y.}~\bibnamefont
  {Li}}, \bibinfo {author} {\bibfnamefont {X.}~\bibnamefont {Chen}}, \ and\
  \bibinfo {author} {\bibfnamefont {M.~P.}\ \bibnamefont {Fisher}},\
  }\href@noop {} {\bibfield  {journal} {\bibinfo  {journal} {Physical Review
  B}\ }\textbf {\bibinfo {volume} {100}},\ \bibinfo {pages} {134306} (\bibinfo
  {year} {2019})}\BibitemShut {NoStop}%
\bibitem [{\citenamefont {Skinner}\ \emph {et~al.}(2019)\citenamefont
  {Skinner}, \citenamefont {Ruhman},\ and\ \citenamefont
  {Nahum}}]{SkinnerNahum2018measure}%
  \BibitemOpen
  \bibfield  {author} {\bibinfo {author} {\bibfnamefont {B.}~\bibnamefont
  {Skinner}}, \bibinfo {author} {\bibfnamefont {J.}~\bibnamefont {Ruhman}}, \
  and\ \bibinfo {author} {\bibfnamefont {A.}~\bibnamefont {Nahum}},\
  }\href@noop {} {\bibfield  {journal} {\bibinfo  {journal} {Physical Review
  X}\ }\textbf {\bibinfo {volume} {9}},\ \bibinfo {pages} {031009} (\bibinfo
  {year} {2019})}\BibitemShut {NoStop}%
\bibitem [{\citenamefont {Houdayer}\ and\ \citenamefont
  {Hartmann}(2004)}]{houdayer2004low}%
  \BibitemOpen
  \bibfield  {author} {\bibinfo {author} {\bibfnamefont {J.}~\bibnamefont
  {Houdayer}}\ and\ \bibinfo {author} {\bibfnamefont {A.~K.}\ \bibnamefont
  {Hartmann}},\ }\href@noop {} {\bibfield  {journal} {\bibinfo  {journal}
  {Physical Review B}\ }\textbf {\bibinfo {volume} {70}},\ \bibinfo {pages}
  {014418} (\bibinfo {year} {2004})}\BibitemShut {NoStop}%
\bibitem [{\citenamefont {Kawashima}\ and\ \citenamefont
  {Ito}(1993)}]{kawashima1993critical}%
  \BibitemOpen
  \bibfield  {author} {\bibinfo {author} {\bibfnamefont {N.}~\bibnamefont
  {Kawashima}}\ and\ \bibinfo {author} {\bibfnamefont {N.}~\bibnamefont
  {Ito}},\ }\href@noop {} {\bibfield  {journal} {\bibinfo  {journal} {Journal
  of the Physical Society of Japan}\ }\textbf {\bibinfo {volume} {62}},\
  \bibinfo {pages} {435} (\bibinfo {year} {1993})}\BibitemShut {NoStop}%
\bibitem [{\citenamefont {Gullans}\ and\ \citenamefont
  {Huse}(2019)}]{gullans2019dynamical}%
  \BibitemOpen
  \bibfield  {author} {\bibinfo {author} {\bibfnamefont {M.~J.}\ \bibnamefont
  {Gullans}}\ and\ \bibinfo {author} {\bibfnamefont {D.~A.}\ \bibnamefont
  {Huse}},\ }\href@noop {} {\bibfield  {journal} {\bibinfo  {journal} {arXiv
  preprint arXiv:1905.05195}\ } (\bibinfo {year} {2019})}\BibitemShut {NoStop}%
\bibitem [{\citenamefont {Bao}\ \emph {et~al.}(2020)\citenamefont {Bao},
  \citenamefont {Choi},\ and\ \citenamefont {Altman}}]{bao2019theory}%
  \BibitemOpen
  \bibfield  {author} {\bibinfo {author} {\bibfnamefont {Y.}~\bibnamefont
  {Bao}}, \bibinfo {author} {\bibfnamefont {S.}~\bibnamefont {Choi}}, \ and\
  \bibinfo {author} {\bibfnamefont {E.}~\bibnamefont {Altman}},\ }\href@noop {}
  {\bibfield  {journal} {\bibinfo  {journal} {Physical Review B}\ }\textbf
  {\bibinfo {volume} {101}},\ \bibinfo {pages} {104301} (\bibinfo {year}
  {2020})}\BibitemShut {NoStop}%
\bibitem [{\citenamefont {Jian}\ \emph {et~al.}(2020)\citenamefont {Jian},
  \citenamefont {You}, \citenamefont {Vasseur},\ and\ \citenamefont
  {Ludwig}}]{jian2019measurement}%
  \BibitemOpen
  \bibfield  {author} {\bibinfo {author} {\bibfnamefont {C.-M.}\ \bibnamefont
  {Jian}}, \bibinfo {author} {\bibfnamefont {Y.-Z.}\ \bibnamefont {You}},
  \bibinfo {author} {\bibfnamefont {R.}~\bibnamefont {Vasseur}}, \ and\
  \bibinfo {author} {\bibfnamefont {A.~W.}\ \bibnamefont {Ludwig}},\
  }\href@noop {} {\bibfield  {journal} {\bibinfo  {journal} {Physical Review
  B}\ }\textbf {\bibinfo {volume} {101}},\ \bibinfo {pages} {104302} (\bibinfo
  {year} {2020})}\BibitemShut {NoStop}%
\bibitem [{\citenamefont {Preskill}(2018)}]{PreskillNotes}%
  \BibitemOpen
  \bibfield  {author} {\bibinfo {author} {\bibfnamefont {J.}~\bibnamefont
  {Preskill}},\ }\href@noop {} {\enquote {\bibinfo {title} {Lecture notes for
  physics 219: Quantum computation},}\ } (\bibinfo {year} {2018})\BibitemShut
  {NoStop}%
\bibitem [{\citenamefont {Wilde}(2013)}]{wilde2013quantum}%
  \BibitemOpen
  \bibfield  {author} {\bibinfo {author} {\bibfnamefont {M.~M.}\ \bibnamefont
  {Wilde}},\ }\href@noop {} {\emph {\bibinfo {title} {Quantum information
  theory}}}\ (\bibinfo  {publisher} {Cambridge University Press},\ \bibinfo
  {year} {2013})\BibitemShut {NoStop}%
\bibitem [{\citenamefont {Devetak}\ and\ \citenamefont
  {Shor}(2005)}]{devetak2005capacity}%
  \BibitemOpen
  \bibfield  {author} {\bibinfo {author} {\bibfnamefont {I.}~\bibnamefont
  {Devetak}}\ and\ \bibinfo {author} {\bibfnamefont {P.~W.}\ \bibnamefont
  {Shor}},\ }\href@noop {} {\bibfield  {journal} {\bibinfo  {journal}
  {Communications in Mathematical Physics}\ }\textbf {\bibinfo {volume}
  {256}},\ \bibinfo {pages} {287} (\bibinfo {year} {2005})}\BibitemShut
  {NoStop}%
\bibitem [{\citenamefont {Collins}(2003)}]{collins2003moments}%
  \BibitemOpen
  \bibfield  {author} {\bibinfo {author} {\bibfnamefont {B.}~\bibnamefont
  {Collins}},\ }\href@noop {} {\bibfield  {journal} {\bibinfo  {journal}
  {International Mathematics Research Notices}\ }\textbf {\bibinfo {volume}
  {2003}},\ \bibinfo {pages} {953} (\bibinfo {year} {2003})}\BibitemShut
  {NoStop}%
\end{thebibliography}%

\end{document}

\title{Supplementary Online Material for ``Quantum Error Correction in Scrambling Dynamics and Measurement-Induced Phase Transition''}
\date{\today}

\author{Soonwon Choi}
\thanks{SC and YB contributed equally to this work.}
\affiliation{Department of Physics, University of California Berkeley, Berkeley, CA 94720, USA}
\author{Yimu Bao}
\thanks{SC and YB contributed equally to this work.}
\affiliation{Department of Physics, University of California Berkeley, Berkeley, CA 94720, USA}
\author{Xiao-Liang Qi}
\affiliation{Stanford Institute for Theoretical Physics, Stanford University, Stanford, CA 94305, USA}
\author{Ehud Altman}
\affiliation{Department of Physics, University of California Berkeley, Berkeley, CA 94720, USA}
\affiliation{Materials Science Division, Lawrence Berkeley National Laboratory, Berkeley, CA 94720, USA}

\maketitle

\tableofcontents

\section{Random Clifford  circuits as unitary 2-design}\label{sec:fp}

One of the main results of our work relies on the decoupling inequality, which requires that  the set of random unitaries to be averaged over forms a unitary 2-design.
While it is well-known that the $n$-qubit Clifford group forms a unitary $2$-design~\cite{divincenzo2002quantum}, it is yet to be verified that an ensemble of random quantum circuits of depth $d$ made out of 2-qubit Clifford gates also approximates a unitary 2-design for $n$ qubits.
In this section, we numerically compute the frame potential for such unitary circuits, which quantifies the extent to which it approximates unitary designs. Our results confirm that the ensemble of depth $d$ circuits of local 2-qubit Clifford gates indeed approximates a unitary 2-design when $d$ is large. In what follows, we 
first review the frame potential, introduce our algorithm to compute it, and then present numerical results.

\subsection{Frame potential}
The $k$-th frame potential of a unitary ensemble $\nu$ is defined by
\begin{equation}\label{eqn:fp}
F_\nu^{(k)} = \frac{1}{\abs{\nu}^2}\sum_{U, V \in \nu}\abs{\tr\left(U^\dagger V\right)}^{2k},
\end{equation}
where $|\nu|$ denotes the order of the ensemble.
One of the nice properties of the $k$-th frame potential $F_\nu^{(k)}$ is that this quantity is minimized when the unitary ensemble is drawn from the Haar measure (in which the summation in Eq.~\eqref{eqn:fp} is replaced by integration):
\begin{equation}
F_{\nu}^{(k)} \geq F_{\mu_{haar}}^{(k)} = k!.
\end{equation}
Furthermore, it is well known that the $k$-th frame potential saturates this lower bound if and only if the unitary ensemble $\nu$ forms a unitary $k$-design~\cite{roberts2017chaos}.
Therefore, by explicitly computing the second frame potential $F_\nu^{(2)}$, we can verify if the ensemble of Clifford circuits forms an approximate $2$-design.

\subsection{Numerical algorithm}
\begin{figure}[t!]
	\includegraphics[width=0.8\textwidth]{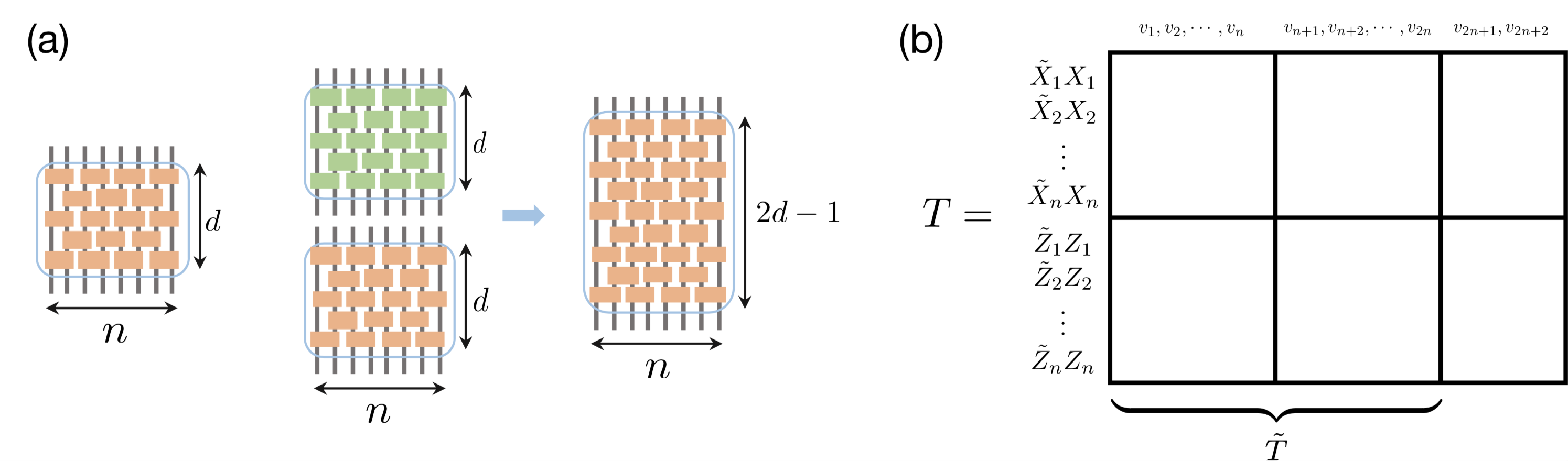}
	\caption{(a) Left: the layout of the random Clifford circuit $U$ considered in this work. For a pair of circuits $U$ and $V$ drawn from an ensemble  $\nu_d$ of depth $d$ circuits, $U^\dagger V$ is, equivalently, a unitary drawn from $\nu_{2d-1}$ of depth $2d-1$.
		(b) An illustration of matrix $T$. 
		The first and the second $n$ rows of $T$ represent operators $\tilde{X}_1X_1, \cdots, \tilde{X}_nX_n$ and $\tilde{Z}_1Z_1, \cdots, \tilde{Z}_nZ_n$, respectively.
		Each operator is specified by using a length $2n + 2$ binary vector, based on the Eq.~\eqref{eq:pauli}.}
	\label{suppfig:FP_Algo}
\end{figure}

While it is well known that the unitary evolution of a quantum state under Clifford gates can be efficiently simulated using classical computers~\cite{bennett1996mixed, calderbank1997quantum, gottesman1996class, gottesman1998heisenberg}, this does not imply that one can explicitly evaluate the matrix that implements the unitary time evolution.
Computing the exact frame potential for an ensemble of Clifford circuits is a formidable task mainly due to two reasons.
First, in a na\"ive approach, taking the trace of a Clifford unitary requires simulating the evolution of exponentially many different initial states owing to the large Hilbert space.
Second, the summation over every element, $U, V$, from the ensemble of random Clifford circuit is computationally expensive due to the large size of the  ensemble.
While the second difficulty can be resolved by performing Monte Carlo sampling of the group elements, the first challenge is non-trivial.

In this section, we provide an efficient algorithm to compute the trace of a Clifford circuit.
In comparison to the na\"ive approach of simulating the evolution of exponentially many different initial states, our approach takes only a polynomial time as a function of system size $n$.

Recall that the $k$-th frame potential for an ensemble of Clifford circuits of depth $d$ is of the form
\begin{equation}
F_{\nu_d}^{(k)} = \frac{1}{\abs{\nu_d}^2}\sum_{U, V \in \nu_d}\abs{\tr(U^\dagger V)}^{2k}.
\end{equation}
Here, we focus on the layout of the circuit shown in Fig.~\ref{suppfig:FP_Algo}(a), while our technique introduced here is more broadly applicable.
Since both $U$ and $V$ consist of $d$ layers of random Clifford gates, we note that $U^\dagger V$ can be also regarded as a circuit of depth $2d-1$ without loss of generality.
Therefore, we can rewrite the frame potential as
\begin{equation}\label{eq:fp_circuit}
F_{\nu_d}^{(k)} = \frac{1}{\abs{\nu_{2d-1}}}\sum_{U \in \nu_{2d-1}}\abs{\tr U}^{2k} 
=\frac{1}{\abs{\nu_{2d-1}}}\sum_{U \in \nu_{2d-1}}\abs{\tr U \tr U^\dagger}^{k} 
=\frac{1}{\abs{\nu_{2d-1}}}\sum_{U \in \nu_{2d-1}}\abs{Q_U}^{k},
\end{equation}
where $Q_U \equiv \tr U \tr U^\dagger$. 
Thus, our problem reduces to computing $Q_U$ exactly for a given unitary $U$, and then performing a Monte-Carlo sampling of $U$ over different realizations of the depth $2d-1$ random Clifford circuits. Below, we focus on the computation of $Q_U$ assuming $U$ is given as a Clifford circuit.

Our key idea is to further simplify the expression $Q_U$ using Pauli operators.
In particular, we are interested in evaluating $Q_U$ without having to explicitly construct the unitary matrix $U$.
This is in fact possible, because, by the definition of the Clifford group, the unitary $U$ can be fully characterized by specifying how generators of $n$-qubit Pauli group $\mathcal{P}$ transform under the conjugation by $U$~\cite{aaronson2004improved}.
To this end, it is important to introduce an efficient notation to denote a Pauli element.
Here and below, we adapt and extend the notation for Pauli operators in the existing literature~\cite{aaronson2004improved} and denote an element in the $n$-qubit Pauli group by a binary string $v = v_1 v_2 \cdots v_{2n+2}$ of length $2n + 2$:
\begin{equation}\label{eq:pauli}
P_v =  (-1)^{v_{2n+1}} i^{v_{2n+2}} \prod_{j=1}^n K_j(v_j, v_{j+n}),
\end{equation}
where $K_j (1,0) = X_j$, $K_j (0,1) = Z_j$, or $K_j (1,1) = Y_j$ represents one of the Pauli operators for a qubit at site $j$, following the convention in \cite{aaronson2004improved}.
In other words, the first $2n$ bits in a string $v$ specify the $n$-qubit Pauli string, and the last two digits control the overall coefficient.
We will find that the global prefactor $(-1)^{v_{2n+1}} i^{v_{2n+2}}$ is often not very important other than that it gives rise to exactly four elements in $\mathcal{P}$ per a single Pauli string (with different prefactors).
Hence, for notational brevity, we denote the first $2n$ bits of $v$ as $\bar{v} = v_1 v_2 \dots v_{2n}$.
In this way, the product of two Pauli strings can be concisely represented as a simple XOR operation on corresponding binary strings: if $P_v \equiv P_u P_w$, then $\bar{v} = \bar{u} + \bar{w}$, where the ``$+$'' operator should be interpreted as element-wise XOR operations.
This establishes the one-to-one correspondence between a binary string of length $2n$ and every group element in $\mathcal{P}$ up to a prefactor. In particular, the group multiplication in $\mathcal{P}$ corresponds to a linear operation in the binary string. Furthermore, the set of binary strings forms a vector space with respect to XOR operations. 
We note that the prefactor of $v$ can be also computed from $u$ and $w$.

In order to compute $Q_U$, we first re-express it using the fact that the Pauli group $\mathcal{P}$ forms a unitary $1$-design. More specifically, for any $n$-qubit operator $O$, we have, $ \tr(O) \cdot \mathbbm{1} =  \frac{1}{4D} \sum_{P\in\mathcal{P}} P O P^\dagger$. Then, it follows:
\begin{equation}\label{eq:trU2}
Q_U = 
\tr U^\dagger \tr U = \tr(U^\dagger \mathbbm{1} \cdot \tr(U)) = \frac{1}{D} \sum_{P \in \mathcal{P}^+} \tr(U^\dagger P U P^\dagger) = \frac{1}{D} \sum_{P \in \mathcal{P}^+} \tr(\tilde{P} P^\dagger),
\end{equation}
where $D=2^n$ is the Hilbert space dimension, $\tilde{P} \equiv U^\dagger P U$, and the summation is over the operators in the set  $\mathcal{P}^+ = \{\mathbbm{1}, X, Y, Z\}^{\otimes n}$, ignoring the irrelevant complex prefactor (which cancels the factor $1/4$).
We note that $\tilde{P}P^\dagger$ is also a Pauli operator and its trace is non-vanishing if and only if $\tilde{P}P^\dagger \propto \mathbbm{1}$. Therefore, in order to evaluate $Q_U$, we only need to count how many $P\in \mathcal{P}^+$ gives rise to a non-vanishing contribution. We denote the set of such Pauli operators as $\mathcal{K}_U$.

Our key observation is that $\mathcal{K}_U$ is closed under multiplications to form a subgroup of $\mathcal{P}$, and its binary representation form a linear vector space over binary field. Therefore, counting the number of element in $\mathcal{K}_U$ can be efficiently achieved by computing the dimension of the vector space $\mathcal{K}_U$. Here and below, we use the same notation $\mathcal{K}_U$ to refer to both the vector space and the subgroup of Pauli group whenever there is no ambiguity.
Below, we will discuss some important properties of $\mathcal{K}_U$.
In particular, we will demonstrate that the computation of  $Q_U$ always falls into one of three cases:
\begin{itemize}
	\item $\mathcal{K}_U$ contains only the identity operator. Namely, the Pauli operator $P$ satisfies $\tilde{P}P^\dagger \propto \mathbbm{1}$ only when $P = \mathbbm{1}$. $Q_U = 1$;
	\item $\mathcal{K}_U$ of dimension $N$ is generated by $N$  Pauli operators $P$ with $\tilde{P}P^\dagger = \mathbbm{1}$. $\mathcal{K}_U$ contains $2^N$ Pauli operators, $Q_U = 2^N$;
	\item There exists at least one generator $P$ of $\mathcal{K}_U$ satisfying $\tilde{P}P^\dagger = -\mathbbm{1}$. 
	Then, there are equal numbers of Pauli operators $P$ with $\tilde{P}P^\dagger = \mathbbm{1}$ and $P$ with $\tilde{P}P^\dagger = \mathbbm{-1}$, i.e., the operators satisfying $\tilde{P}P^\dagger = \mathbbm{1}$ and $\mathbbm{-1}$ come in pairs. $Q_U = 0$.
\end{itemize}

To see this, we notice that, for any $P_v \in \mathcal{P}$, $\tr(\tilde{P_v}P_v^\dagger)$ is always real since
\begin{equation}
\tr(\tilde{P_v}P_v^\dagger)^* = \tr(U^\dagger P_v U P_v^\dagger)^* =  \tr(P_v U^\dagger P_v^\dagger U) = (-1)^{2v_{2n+2}} \tr(P_v^\dagger U^\dagger P_v U) = \tr(\tilde{P_v}P_v^\dagger).
\end{equation}
This implies $\tr(\tilde{P_v}P_v^\dagger) \neq 0$ if and only if $\tilde{P_v}P_v^\dagger = \pm \mathbbm{1}$. We use $\mathcal{K}_U^\pm$ to denote the set of operators $P_v$ satisfying $\tilde{P_v}P_v^\dagger = \pm \mathbbm{1}$, respectively. $\mathcal{K}_U^+$ forms a normal subgroup of $K_U$. If $\mathcal{K}_U^-$ is trivial, i.e., contains no element, $\mathcal{K}_U^+$ is exactly $\mathcal{K}_U$. When $\mathcal{K}_U^-$ is non-trivial, we require at least one generator $P$ of $\mathcal{K}_U$ satisfies $\tilde{P}P^\dagger = - \mathbbm{1}$. In this case, $\mathcal{K}_U^+$ becomes the maximal normal subgroup of $\mathcal{K}_U$, and the quotient group $\mathcal{K}_U/\mathcal{K}_U^+ = \mathbb{Z}/2\mathbb{Z}$. Therefore,  $\mathcal{K}_U^\pm$ contain the same number of elements, and, as a result, $Q_U = 0$. 

Due to this correspondence between $P$ satisfying the property $\tilde{P} P^\dagger \propto \mathbbm{1}$, and the linear space $\mathcal{K}_U$, counting the number of such operators can be efficiently accomplished by finding out a set of mutually commuting generators of the group, or, equivalently, the basis of the linear space $\mathcal{K}_U$ in the binary representation. Namely, the number of elements in $\mathcal{K}_U$ is given by $2^{m_\mathcal{K}}$, where $m_\mathcal{K}$ is the dimension of $\mathcal{K}_U$. In the second case where all operators $P$ in $\mathcal{K}_U$ satisfy $\tilde{P} P^\dagger = \mathbbm{1}$, $Q_U = 2^{m_\mathcal{K}}$.

Now, the calculation of $Q_U$ reduces to figuring out $m_\mathcal{K}$ and the existence or the absence of any generator $P$ satisfying $\tilde{P} P^\dagger = -\mathbbm{1}$. 
In order to find out $m_\mathcal{K}$, we introduce a matrix $T$ of size $2n$ by $2n+2$. The first and the second $n$ rows of $T$ represent the Pauli operators $\tilde{X_i}X_i$ and $\tilde{Z_i}Z_i$, respectively [see Fig.~\ref{suppfig:FP_Algo}(b)]. We denote the first $2n$ columns of $T$ by $\bar{T}$.
Since the identity operator $\mathbbm{1} \in \mathcal{P}$ corresponds to the zero in the binary representation, one can easily check that the linear space $\mathcal{K}_U$ can be determined from the kernel of $\bar{T}$ (over the binary field).
Furthermore, one can determine the sign of $\tilde{P_\alpha}P_\alpha$ for every basis vector in $\mathcal{K}_U$ by explicitly performing the effective ``Gaussian elimination'' (or row operations) that properly accounts for the changes in the prefactors of Pauli operators.
Motivated from Ref.~\cite{aaronson2004improved}, we define a modified \textit{rowsum()} function.
The \textit{rowsum()} function takes two rows of $T$ (corresponding to operators $\tilde{P_u}P_u^\dagger$ and $\tilde{P_v}P_v^\dagger)$ as input and returns a binary representation for the operator $\tilde{P_v}\tilde{P_u}P_u^\dagger P_v^\dagger$:
\begin{equation}
\text{rowsum}(\tilde{P_u}P_u^\dagger, \tilde{P_v}P_v^\dagger) = \tilde{P_v}\tilde{P_u}P_u^\dagger P_v^\dagger = (-1)^{-\bar{t}\Lambda_n\bar{v}}\tilde{P_v}P_v P_t,
\end{equation}
where $P_t = \tilde{P_u}P_u^\dagger$, and $\Lambda_n = \left[0, \mathbbm{1}_n; -\mathbbm{1}_n, 0\right]$ is the symplectic form.
By checking the value in the $(2n+1)$-th column, we can determine the sign for generators $\tilde{P_v}P_v$ of the kernel $\mathcal{K}_U$. In this way, we can compute $Q_U$ for each realization of the circuit and further obtain the frame potential.

\subsection{Numerical results}

\begin{figure}[t!]
	\centering
	\includegraphics[width=\linewidth]{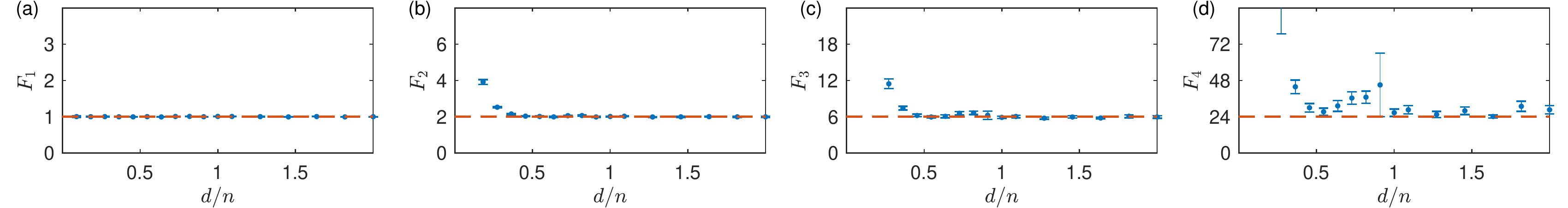}
	\caption{Numerical computation of the first, second, third, and fourth frame potentials (from (a) to (d)) as a function of circuit depth $2\leq d\leq 44$ for system size $n = 22$ qubits. The frame potentials are estimated from $50000$ randomly generated Clifford circuits. The depths of circuit $d$ is taken from $2$ to $44$ (blue markers). Orange dashed lines represent the corresponding values for the Haar random unitary ensemble.
	}
	\label{suppfig:fp}
\end{figure}

The results of our numerical calculations are summarized in Fig.~\ref{suppfig:fp}, which shows that,
the first, second, and third frame potentials for the random Clifford circuit ensemble approach to corresponding values for a unitary $1,2$ and $3$-design when the depth of circuit $d$ is sufficiently large, $\sim O(n)$, as predicted in \cite{brandao2016local}. 
In contrast, the fourth frame potential significantly deviates from the value for a unitary $4$-design, which is expected since it has been proved that even the $n$-qubit Clifford group does not form a unitary $4$-design~\cite{webb2015clifford}.

In the model proposed in the main text, each cluster consists of $m = 11$ qubits, hence a nearest neighboring cluster pair have total $n = 2m = 22$ qubits.
Our numerical results in Fig.~\ref{suppfig:fp}(b) indicate that random Clifford circuits of depth $d = 44$ are sufficient to approximate a unitary $2$-design in such a case.

\section{Detailed numerical simulation results for the entanglement growth}
\begin{figure}[t!]
	\centering
	\includegraphics[width=\linewidth]{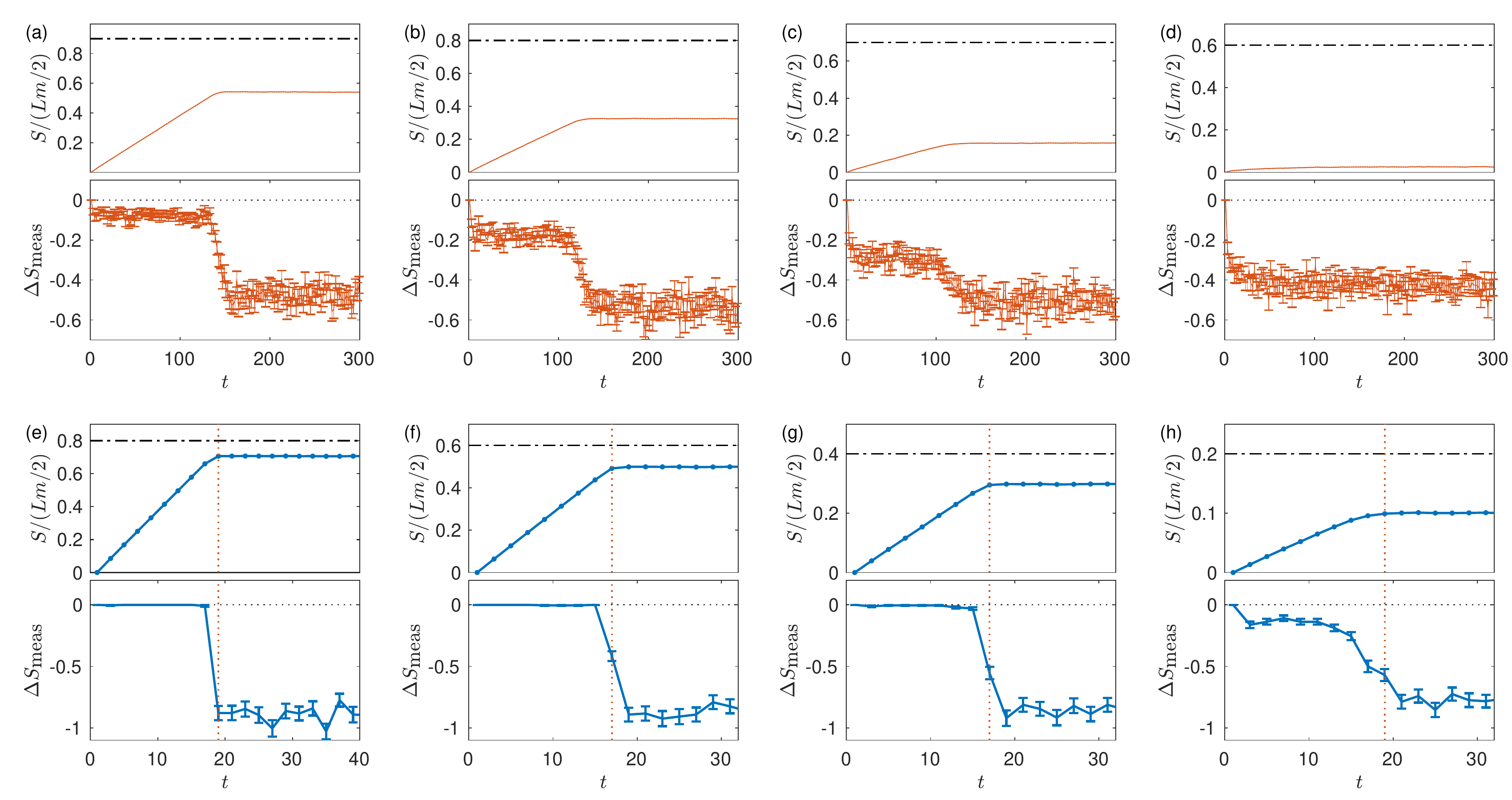}
	\caption{Half-chain entanglement dynamics in various parameter regimes. Simulations are performed for systems with  $L = 32$, $m = 11$. 
		Upper panels in (a-h): the growth of entanglement entropy per qubit as a function of time $t$.
		Black dash-dotted lines indicate $1-p$, which correspond to the maximum possible entanglement entropy per qubit after projective measurements. 
		Lower panels in (a-h): the change of entanglement entropy before and after random projective measurements in each time step. The errorbars represent the standard deviation of entanglement reduction by measurements.
		(a-d): The depth of the local Clifford circuit $d = 3$, and the measurement fraction is $p=0.1, 0.2, 0.3, 0.4$, respectively (from left to right).
		(e-h): The depth of the local Clifford circuit  $d = 44$, and $p=0.2, 0.4, 0.6, 0.8$, respectively (from left to right).
		In this regime, the local circuit approximates a unitary $2$-design.
		The red dotted vertical lines in (e) and (f) indicate when the entanglement entropy per qubit, $S/(Lm/2)$, reaches its steady-state value.
		All the results in this figure are averaged over $240$ different realizations of the random unitary circuit.
	}
	\label{suppfig:growth}
\end{figure}
Figure~\ref{suppfig:growth} provides detailed information on the entanglement growth and saturation in vaious parameter regimes.
We focus on two different values of the local circuit depth, $d = 3$ or $44$, and simulate the quantum dynamics for various values of the measurement fraction $p$~\footnote{We measure $pm$ number of qubits within each block after applying a layer of unitary gates to pairs of qubit blocks. For a noninteger $pm$, the number of measured  qubits is determined from a binomial distribution between $\lfloor pm \rfloor$ and $\lceil pm \rceil$ with the mean value being $pm$.}.
When $d = 3$, each local circuit fails to approximate a unitary $2$-design, and quantum information cannot be fully scrambled even within the local Hilbert space of a $m$-qubit box.
In this regime, the growth of entanglement entropy can be significantly affected by random projective measurement.
Indeed, we find that the projective measurements reduce the half-chain entanglement starting from the early time evolution at $t=0$.

In contrast, when $d = 44$, each Clifford unitary $U$ approximates a unitary $2$-design for $2m = 22$ qubits as demonstrated in the previous section. One can apply the decoupling inequality as discussed in the main text (see Section~\ref{sec:decoupling_inequality} for its derivation).
In this regime, the entanglement entropy should not be significantly decreased by projective measurements during early time evolution as long as $\gamma < 1 - p$, or more precisely $2^{-(1-\gamma - p) m}\ll 1$, where $\gamma$ is the entanglement entropy per qubit.
This regime is indicated by using a vertical line with the identification $\gamma = S/(Lm/2)$.
Strictly speaking, this identification is not exact, since the relevant $\gamma$ for decoupling theorem should have been obtained from the entanglement entropy between a neighboring qubit blocks and the rest of the system, rather than from the half-chain entanglement $S$.
Still, we expect the qualitatively similar behavior.
Our expectation is explicitly verified in Figs.~\ref{suppfig:growth}(e-g), where $\Delta S_{meas} \approx 0$ within errorbars.
For the largest measurement probability $p=0.8$ in Fig.~\ref{suppfig:growth}(h), $\Delta S_{meas}$ is nonnegligible as the required condition, $2^{-(1-\gamma - p) m}\ll 1$, for our improve decoupling inequality in Section~\ref{sec:decoupling_inequality} is no longer satisfied.

\section{Detailed numerical simulation results for the phase transition}

In this section, we present detailed numerical simulation results for the entanglement phase transition. 
We first study the phase transition when the size of qubit block $m = 11$ is large. We extract critical measurement probabilities $p_c$ and critical exponents $\nu$ using the finite-size scaling analysis of half-chain entanglement entropy and tripartite mutual information in Sec.~\ref{sec:half-chain} and~\ref{sec:tri_mul}, respectively. 
Furthermore, we investigate the phase transition for various $m$ with a fixed $d/m = 3$ in Sec.~\ref{sec:transition_m}. We extract $p_c$, $\nu$ as well as the prefactor of logarithmic scaling of entanglement entropy at the critical point for various $m$ and compare the results to the existing theoretical predictions.

\subsection{Half-chain entanglement entropy}\label{sec:half-chain}

\begin{figure}[t!]
	\centering
	\includegraphics[width=\linewidth]{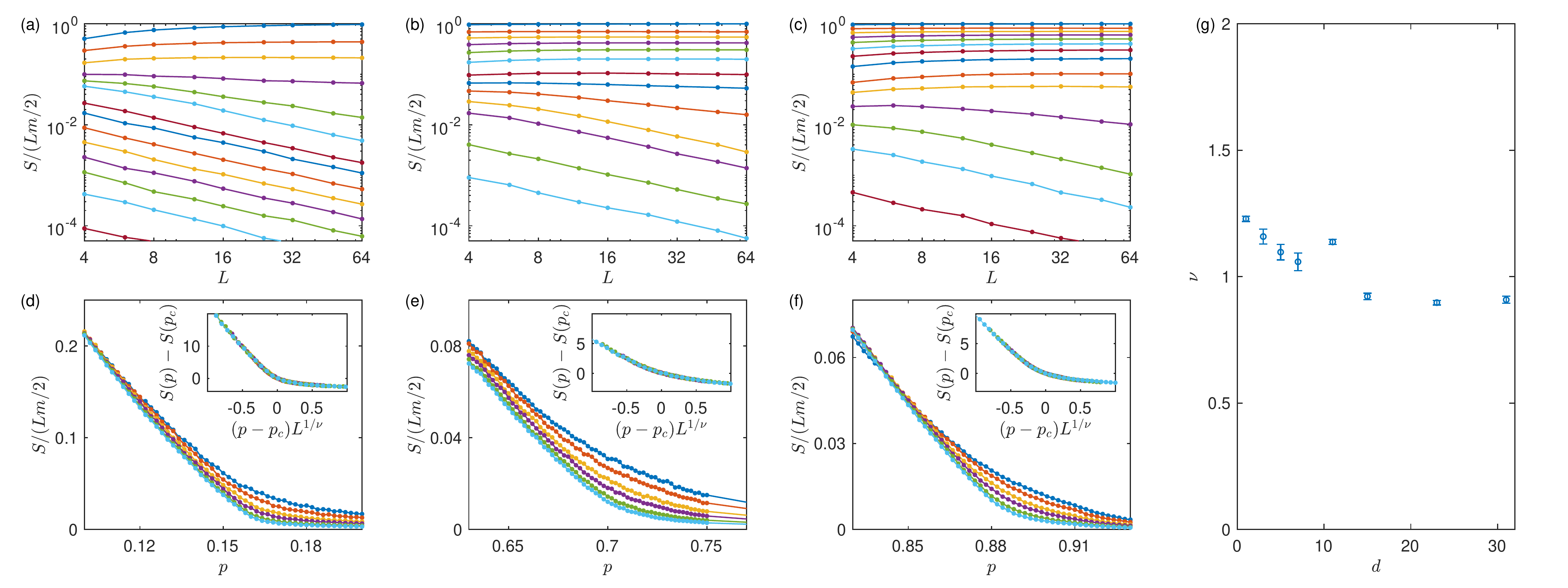}
	\caption{
		Finite size scaling analysis and critical exponents. 
		(a-c) The entanglement density $S/(Lm/2)$ as a function of system size $L$ for different depths of local circuits, $d = 1, 7, 31$, respectively (from left to right).
		Different curves correspond to our numerical results with different measurement fractions.
		System size $L$ ranges from $4$ to $64$.
		(d-f) The entanglement density as a function of measurement fraction for $d=1, 7, 31$, respectively (from left to right).
		Different curves correspond to results from different system sizes $L = 12, 16, 24, 32, 48, 64$. Data collapses are presented in the insets using the scaling hypothesis Eq. \ref{eq:scaling_new}.
		(g) Numerically extracted critical exponents $\nu$ for different values of $d$. 
		The error bars are estimated using the bootstrapping method.
	}
	\label{suppfig:critexpo_new}
\end{figure}

Following the entanglement scaling hypothesis proposed in a recent work~\cite{li2019measurement},
we perform a finite-size scaling analysis with the scaling ansatz:
\begin{equation}\label{eq:scaling_new}
S(p, L) = \alpha \ln L + \mathcal{F}\left( (p-p_c)L^{1/\nu} \right),
\end{equation}
where $\alpha$ is a constant, characterizing the logarithmic entropy at the critical point, $p_c$ is the critical measurement fraction,  $\nu$ is the correlation length critical exponent, and $\mathcal{F}(x)$ is a universal function.
We expect that the universal function $\mathcal{F}\left(x\right)$ takes the following qualitative behaviors:
\begin{equation}
\mathcal{F}\left( x \right) \approx \left\{ \begin{array}{cc}
\abs{x}^\nu & (x \rightarrow -\infty) \\
\mathrm{const.} & (x = 0) \\
-\alpha \nu \ln x & (x \rightarrow \infty).
\end{array}\right.
\end{equation}
Thus, in the thermodynamic limit $L \rightarrow \infty$, when $p>p_c$, $S(p, L)$ converges to a constant with no dependence on $L$, indicating the area-law phase. In the case $p<p_c$, $S(p, L)$ scales linearly in $L$ with a log correction.
In practice, we substract the entropy at the critical point, $\alpha\ln L$, (with numerically optimized $p_c$) from both sides of Eq. (\ref{eq:scaling_new}), converting it into a conventional finite size scaling form~\cite{SkinnerNahum2018measure}, i.e.,
\begin{equation}
S(p, L) - S(p_c, L) = \mathcal{F}\left( (p-p_c)L^{1/\nu} \right).\label{eqn:scaling_entropy}
\end{equation}
More specifically, we numerically optimize  the parameter $p_c$ and $\nu$ by minimizing the cost function 
\begin{equation}
Q = \frac{1}{\mathcal{N}}\sum_{i,j}\frac{(y_{ij} - Y_{ij})^2}{dy_{ij}^2 + dY_{ij}^2},
\end{equation}
where $Y_{ij}$, $dY_{ij}$ are the values given by the scaled function and its standard error at $x_{ij}$, and $y_{ij}$, $dy_{ij}$ are the data points at $x_{ij}$. The index $j$ labels different system sizes and $i$ labels different measurement fractions. The detailed algorithm can be found in \cite{houdayer2004low, kawashima1993critical}.

In Figs.~\ref{suppfig:critexpo_new}(a-f), we present our numerical results for the phase transition with various local circuit depths $d$ and fixed qubit block size $m = 11$.
The half-chain entanglement entropy normalized by the number of qubits $S/(Lm/2)$ approaches a constant in the volume-law phase, while it decays as $1/L$ in the area-law phase [see Figs.~\ref{suppfig:critexpo_new}(a-c)].
$S/(Lm/2)$ as a function of measurement probability $p$ shows that the transition becomes sharper when increasing the system size [see Figs.~\ref{suppfig:critexpo_new}(d-f)].
We obtain data collapses in the insets of Figs.~\ref{suppfig:critexpo_new}(d-f) using the scaling formula in Eq.~\ref{eqn:scaling_entropy}~\cite{SkinnerNahum2018measure}.
Critical measurement probabilities $p_c$ and critical exponents $\nu$ are extracted by optimizing the cost function $Q$ in two steps: (1) choose a $p_c$ in the critical regime and find the minimum of the cost function $Q_{\text{min}}(p_c)$ for the given $p_c$ by optimizing over $\nu$; (2) find the global minimum of $Q_{\text{min}}(p_c)$ to extract the optimal $p_c$ and $\nu$.
The extracted
$p_c$ and $\nu$ for various $d$ with fixed $m = 11$ are presented in Fig.~2(e) in the main text and Fig.~\ref{suppfig:critexpo_new}(g), respectively.
We estimate the error bars for $\nu$ in Fig.~\ref{suppfig:critexpo_new}(g) using the bootstrapping method.
More specifically, out of $100$ measurement probabilities in the critical regime for each $d$, we randomly choose $80$ data points and perform the scaling analysis described above to extract $\nu$ and $p_c$. We repeat the analysis for $100$ times and use the standard deviation of $\nu$ as an estimation of the error bars.
We note that the estimated error bars only reflect the goodness of data collapse using the aforementioned method (statistical errors), and do not reflect the accuracy of this method, i.e. any potential systematic errors may not be accounted for.
Using this method, we numerically extract critical exponents $\nu \approx 1$, which is in rough agreement with the results in Refs.~\cite{li2019measurement}. 
However, the extracted $\nu$ fluctuates as a function of $d$.
This is due to the presence of a logarithmic correction to entanglement entropy in the volume-law phase.
The resulting scaling formula in Eq.~\ref{eqn:scaling_entropy} requires an additional optimization over $p_c$, which limits the accuracy $(p_c, \nu)$ of finite size scaling.
For these reasons, we find that extracting $\nu$ from tripartite mutual information is more reliable as we discuss in the next section.

\subsection{Tripartite mutual information}\label{sec:tri_mul}
\begin{figure}[t!]
	\centering
	\includegraphics[width=\linewidth]{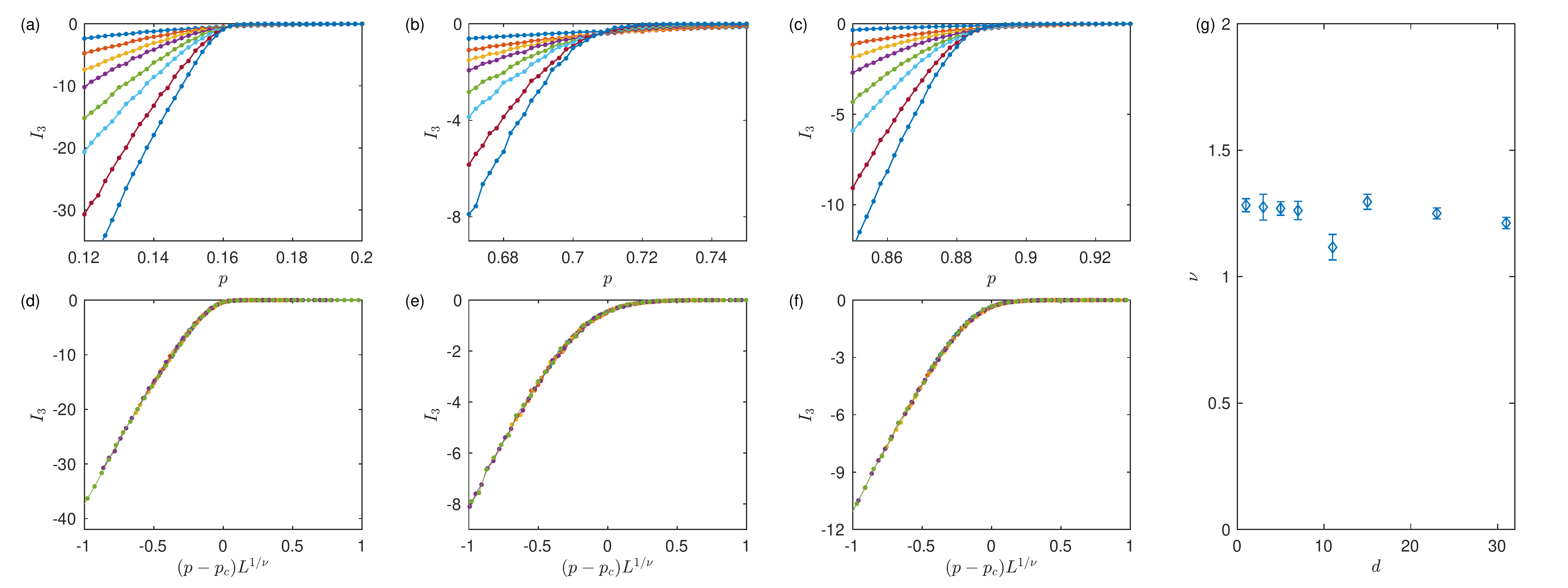}
	\caption{
		Finite size scaling analysis and critical exponents. 
		(a-c) The tripartite mutual information $I_3$ as a function of measurement probability $p$ for different depths of local circuits, $d = 1, 7, 31$, respectively (from left to right).
		Different curves correspond to our numerical results for various system sizes ranging from $4$ to $64$.
		(d-f) Data collapses for $d = 1, 7, 31$ (from left to right) obtained using the scaling hypothesis Eq.~\eqref{eqn:tri_mul_scaling}.
		(g) Extracted critical exponents $\nu$ for various local circuit depth $d$. The numerical values of $p_c$ and $\nu$ are also summarized in Table~\ref{tab:critexpo_tri_mul}.
	}
	\label{suppfig:critexpo_tri_mul}
\end{figure}

In order to improve the accuracy of the extracted critical measurement strength $p_c$ and critical exponent $\nu$, Ref.~\cite{gullans2019dynamical} proposed to use the tripartite mutual information $I_3$ as an alternative probe of the phase transition. More specifically, we consider an one-dimensional chain of qubits with the periodic boundary condition that contains four contiguous subsystems $A$, $B$, $C$ and $D$ of size $L/4$. 
The tripartite mutual information $I_3$ characterizes the nonlocal entanglement among four partitions and is defined as
\begin{align}
    I_3(A:B:C) \equiv S_A + S_B + S_C - S_{AB} - S_{BC} - S_{AC} + S_{ABC},
\end{align}
where $S$ represents the entropy for the corresponding reduced density matrix. This quantity scales with the size of the system $L$ in the volume-law phase, while exhibits an area-law scaling in the area-law phase. More importantly, the logarithmic corrections to the entropy in the volume-law phase cancel in the expression for $I_3(A:B:C)$, and $I_3$ takes an $O(1)$ value at the critical point.

Near the critical point, $I_3$ follows the scaling ansatz~\cite{gullans2019dynamical}:
\begin{align}\label{eqn:tri_mul_scaling}
    I_3(p, L) = \mathcal{G}\left( (p-p_c)L^{1/\nu} \right),
\end{align}
where $\mathcal{G}(\cdot)$ is a universal function. Our finite-size scaling and numerically extracted $p_c$ and $\nu$ are summarized in Fig.~\ref{suppfig:critexpo_tri_mul} and Table~\ref{tab:critexpo_tri_mul}.
The extracted critical exponent $\nu \approx 1.25$ shows no (or very weak) dependence on the depth $d$, suggesting its universal behavior.
The exponent $\nu$ in our model is consistent with the value from the brick-layer random Clifford circuit model studied in Refs.~\cite{li2019measurement}, which (approximately) corresponds to the special case $d = 1$ in our model.

\begin{table}[t!]
	\centering
	\begin{tabular}{|c|cccccccc|}
		\hhline{|=|========|}
		d & 1 & 3 & 5 & 7 & 11 & 15 & 23 & 31 \\
		\hline
		$\nu$ & $1.28\pm 0.03$ & $1.28 \pm 0.05$ & $1.27 \pm 0.03$ & $1.26 \pm 0.04$ & $1.12 \pm 0.05$ & $1.30\pm 0.03$ & $1.25 \pm 0.02$ & $1.21 \pm 0.02$ \\
		$p_c$ & $0.162 $ & $0.412 $ & $0.589 $ & $0.707$ & $0.826 $ & $0.862 $ & $0.883$ & $0.886 $\\
		\hhline{|=|========|}
	\end{tabular}
	\caption{Critical exponents $\nu$  and phase transition points $p_c$ for our toy model in the main text with different depths $d$ for local random Clifford circuits. The results are extracted from the finite-size scaling analysis of tripartite mutual information $I_3$ according to the scaling ansatz given in Eq.~\eqref{eqn:tri_mul_scaling}. The number of qubits in a cluster $m = 11$. The error bars of the critical exponent $\nu$ are estimated using the bootstrapping method.}
	\label{tab:critexpo_tri_mul}
\end{table}

\subsection{Phase transition for various sizes of qubit block}\label{sec:transition_m}

In this section, we focus on deep local Clifford circuits, i.e., $d/(2m) \gtrsim 1$, in which the local circuit approximates a unitary $2$-design (see Sec.~\ref{sec:fp}).
The entire local unitary circuit of depth $d$ can be considered as a single Clifford gate acting on nearest neighbor qudits with a local Hilbert space dimension $q = 2^m$. 
Therefore, changing the size of qubit blocks in this regime is equivalent to tuning local Hilbert space dimension $q$, and it allows us to study the phase transition for various $q$.
The behavior of the limiting case $q\rightarrow \infty$ (or $m\rightarrow \infty$) has been previously discussed for a brick-layer circuits in Ref.~\cite{bao2019theory,jian2019measurement}.

In our model, the critical measurement probability $p_c$ approaches unity in the limit of $m \to \infty$, as we can show by using the newly developed decoupling inequality in Sec.~\ref{sec:decoupling_inequality}.
We note that this limit in our model is different from the $q\rightarrow \infty$ limit in the brick layer random unitary circuit model studied in Refs.~\cite{bao2019theory,jian2019measurement}.
The projective measurements in our model measures a deterministic fraction of qubits within each qubit block, while the measurements in the brick-layer random circuit projects qudits of dimension $q$ (or, equivalently, the entire qubits in a block) with a certain probability.
This distinction may lead to a different universality of the phase transition, which remains largely unexplored.
We also note that Refs.~\cite{bao2019theory,jian2019measurement} has shown that brick-layer random circuit models with measurements can be mapped to a 2D bond percolation problem on square lattice in the $q\rightarrow \infty$ limit.
However, the mapping to the percolation problem is not applicable to our present model in the same limit.

\begin{figure}[t!]
	\centering
	\includegraphics[width=0.8\linewidth]{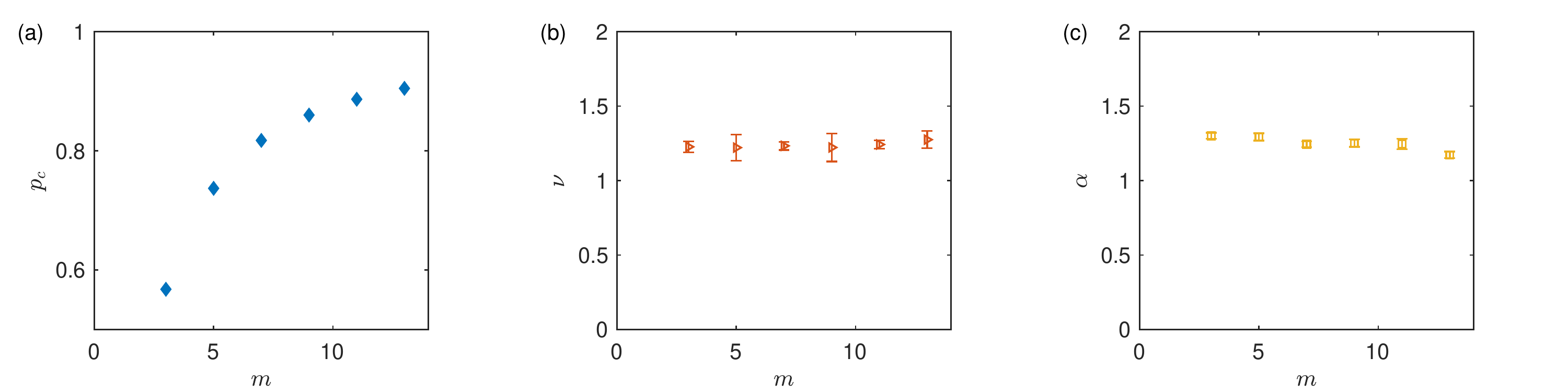}
	\caption{
		Numerically extracted phase transition points and critical exponents for various sizes of qubit clusters $m$ with a fixed relative depth $d/m = 3$, based on the finite size scaling ansatz of $I_3$ in Eq.~\eqref{eqn:tri_mul_scaling} for system sizes up to $L = 64$.
		(a) Critical measurement strength $p_c$ for various $m = 3, 5, 7, 9, 11, 13$.
		(b) Critical exponent $\nu$ for various $m$. The error bars are estimated by using the bootstrapping method.
		(c) Coefficient $\alpha$ for the logarithmic entanglement entropy scaling at the critical point $p_c$ for various $m$.
	}
	\label{suppfig:transition_m}
\end{figure}

Here, we investigate the phase transition as a function of $q$ in our qubit-block model by fixing $d/m = 3$.
We extract critical measurement probabilities $p_c$ and critical exponents $\nu$ for various $m$, as presented in Figs.~\ref{suppfig:transition_m}(a,b). 
The critical measurement probability $p_c$ grows monotonically with $m$ [Fig.~\ref{suppfig:transition_m}(a)].
This is consistent with the prediction that $p_c \to 1$ when $m \to \infty$.
The critical exponent $\nu$ shows no obvious dependence on $m$ and takes a universal value $\nu \approx 1.25$ [Fig.~\ref{suppfig:transition_m}(b)], which suggests the universality of the phase transition remains the same for different $m$'s. 
In addition, the entanglement entropy at the critical point scales logarithmically with the system size:
\begin{align}
    S(p_c, L) = \alpha \log L,
\end{align}
where $\alpha$ is expected to be universal~\cite{SkinnerNahum2018measure,jian2019measurement} and determined by the underlying theory of the critical point.
Here, we evaluate the entropy $S(p_c, L)$ at the critical point for various system sizes up to $L = 64$. We observe a logarithmic scaling of the entanglement entropy and extract $\alpha$ as a function of $m$ [Fig.~\ref{suppfig:transition_m}(c)], which does not show any substantial change for different values of $m$. 

\section{Quantum channel capacity and entanglement phase transition}\label{sec:channel_capacity}
\begin{figure}
    \centering
    \includegraphics[width=0.6\textwidth]{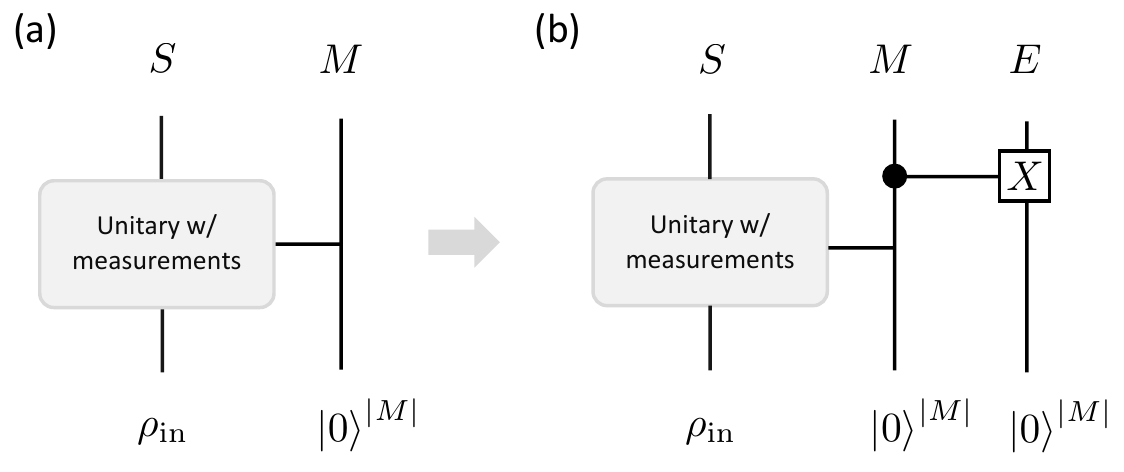}
    \caption{(a) We consider an maximally mixed input state $\rho_\textrm{in}$ evolved under unitary dynamics interspersed with measurements. The output of the system and the classical measurement outcomes are denoted as $S$ and $M$.
    (b) The scenario illustrated (a) can be understood as a unitary evolution by considering the measurements as unitary coupling between the system and measurement device, followed by the dephasing of the measurement device. The measurement device $M$ is dephased through the coupling with an auxiliary environment $E$. The coupling is realized by the controlled-$X$ operation that couples every single ancilla qudit in $M$ to a qudit in $E$. Here, we consider a specific environment $E$ that contains $\abs{M}$ qudits prepared in $\ket{0}$.}
    \label{suppfig:setup}
\end{figure}

In this section, we establish a connection between the quantum channel capacity and the entanglement phase transition.
More specifically, we consider a generic unitary evolution interspersed by projective measurements as a quantum channel jointly acting on the system and measurement devices, and show that the quantum channel capacity $\mathcal{Q}$ of such a channel can be related to the entropy of the system $\langle S \rangle$ conditioned on measurement outcomes when the system is initialized in a certain optimized state [see Fig.~\ref{suppfig:setup}(a)].
We note that the former, $\mathcal{Q}$, characterizes the maximum amount of quantum information that can be transmitted through a quantum channel, while the latter, $\langle S \rangle$ is proposed as an alternative signature to identify the entanglement phase transition in Ref.~\cite{gullans2019dynamical}.
In what follows, we first introduce our setup and notations that would allow us to quantitatively study the quantum channel capacity. Then, we derive the relation between $\mathcal{Q}$ and $\langle S \rangle$.

The quantum channel capacity is given by the coherent information about the input state that remains in the output state, optimized over all possible input encoding state $\rho_\textrm{in}$.
In our case, we are interested in the coherent information stored in the resultant quantum state of the system \emph{as well as} in the set of projective measurement outcomes.
In order to quantify the coherent information that remains in both the system and measurement outcomes, we adapt the weak measurement framework for projective measurements.

In this framework, for each weak measurement, we introduce an ancilla qudit, representing a part of a measurement device $M$, prepared in a predetermined state, e.g. $\ket{0}$.
The ancilla is then correlated with system degrees of freedom by a generic unitary operation.
Finally, the ancilla qudit is dephased by additional degrees of freedom, $E$, that represents environment [Fig.~\ref{suppfig:setup}(b)].
The dephasing by the environment ensures that only the classical information remains in the measurement devices.
By varying the unitary operation that correlates the ancilla and system degrees of freedom, this formulation allows to characterize measurements in arbitrary positive operator-valued measure (POVM). 
The dephasing of $M$ can be realized by coupling to $E$ with generalized controlled-$X$ gates,
\begin{align}
    \mathrm{CX}_{ab} = \ket{0}_a\bra{0}_a \otimes \mathds{1}_b +  \sum_{i=1}^{d-1} \ket{i}_a\bra{i}_a \otimes \exp\left[ -i\frac{\pi}{2}\left( \ket{i}_b\bra{0}_b + \ket{0}_b\bra{i}_b\right)\right],
\end{align}
acting on $a\in M$ (control) and $b\in E$ (target).
Here, $\ket{i}$ with $i\in \{0, 1, \dots d-1\}$ forms the computational basis for a qudit in $M$ or $E$. 
Tracing out every qudit in $E$ after applying the set of $\mathrm{CX}_{ab}$ gates indeed realizes a dephasing channel $\mathcal{D}_{\phi,M}$ applied to $M$.
Therefore, an arbitrary quantum channel $\mathcal{M}$ that consists of generic unitary evolutions interspersed by measurements in any POVM can be formulated as  
\begin{align}
\label{seqn:m}
    \mathcal{M}[\rho] = \mathcal{D}_{\phi,M}\left[U\left( \rho \otimes \left(\ket{0}\bra{0}\right)^{|M|} \right) U^\dagger \right],
\end{align}
with some unitary $U$. Here, $|M|$ denotes the number of qudits in $M$.
Fig.~\ref{suppfig:setup} illustrates our framework, in which each of $M$ and $E$ consists of $|M|$ number of qudits.

In this setting, we investigate the quantum channel capacity of $\mathcal{M}$.
We note that the output of the quantum channel $\mathcal{M}$ can be explicitly divided into two parts: (i) the density matrix of the system quantum state and (ii) classical information (diagonal density matrix) associated with the dephased measurement device $M$.  
The quantum channel capacity $\mathcal{Q}$ for a channel $\mathcal{M}$ is defined by the maximum coherent information per single channel usage when $n$ copies of the channel are simultaneously utilized~\cite{PreskillNotes,wilde2013quantum}:
\begin{align}
\label{seqn:q}
    \mathcal{Q} = \lim_{n \to \infty} \frac{1}{n} \max_{\rho^{(n)}} \mathcal{I}_c\left(\mathcal{M}^{\otimes n}, \rho^{(n)}\right),
\end{align}
where $\rho^{(n)} \in \mathcal{H}^{\otimes n}$ is an input quantum state in the $n$-replicated Hilbert space to be optimized, $\mathcal{I}_c$ is the coherent quantum information defined below.
For a special class of quantum channel, so-called \emph{degradable quantum channels}~\cite{devetak2005capacity,PreskillNotes,wilde2013quantum}, Eq.~\eqref{seqn:q} dramatically simplifies to 
\begin{align}
\label{seqn:single_q}
    \mathcal{Q} =  \max_{\rho} \mathcal{I}_c\left(\mathcal{M}, \rho\right).
\end{align}
This is because, for degradable quantum channels, the quantum channel capacity is additive~\cite{devetak2005capacity,PreskillNotes,wilde2013quantum}.

A degradable quantum channel is defined by the following property.
For a quantum channel $\mathcal{N}$ that transmits a quantum state from a sender $A$ to receiver $A'$, let $\mathcal{N}_c$ be the complementary channel of $\mathcal{N}:\mathcal{H}_A \to \mathcal{H}_{A'}$. That is, we consider an isometric embedding $U_{\mathcal{N}}:\mathcal{H}_A\to \mathcal{H}_{A'}\otimes \mathcal{H}_B$ of $\mathcal{N}$ in an extended Hilbert space, i.e. for any $\rho_A \in \mathcal{H}_A$,
\begin{align}
\mathcal{N}[\rho_A] = \tr_B \left( U_\mathcal{N} \rho_A U_\mathcal{N}^\dagger\right).
\end{align}
Then, the complementary channel $\mathcal{N}_c$ is defined by
\begin{align}
    \mathcal{N}_c[\rho_A] \equiv \tr_{A'} \left( U_\mathcal{N} \rho_A U_\mathcal{N}^\dagger\right).
\end{align}
The channel $\mathcal{N}$ is degradable if there exists another quantum channel $\mathcal{T}$ such that $\mathcal{N}_c = \mathcal{T} \circ \mathcal{N}$.
For any quantum channel of the form in Eq.~\eqref{seqn:m}, its isometric embedding can be written as a unitary.
This is illustrated in Fig.~\ref{suppfig:setup}(a,b).
Therefore, by identifying $A=S$, $A' = SM$ and $B = E$, we find that
\begin{align}
    \mathcal{M}[\rho] &\equiv \tr_E \left( U \rho U^\dagger \right),\\
    \mathcal{M}_c[\rho] &\equiv \tr_{SM} \left( U \rho U^\dagger \right).
\end{align}
Crucially, by tracing out system degrees of freedom, $\mathcal{M}$ can be degraded into $\mathcal{M}_c$, i.e. $\mathcal{M}_c[\rho] \simeq \tr_S \mathcal{M}[\rho]$. This can be easily shown by considering the reduced density matrix of $M$ and $E$ after applying the set of generalized controlled-$X$ gates, which must be of the form $\rho_{ME} = \sum_{ij} \rho_{ij} \ket{i}_M\bra{j}_M \otimes \ket{i}_E \bra{j}_E$. In turn, the reduced density matrices for $M$ and $E$ are given by $\rho_{M(E)} = \sum_{i} \rho_{ii} \ket{i} \bra{i}$ in their corresponding Hilbert spaces.
Below, we focus on establishing a relation between Eq.~\eqref{seqn:single_q} and $\langle S \rangle$.

For an input state $\rho$ and a quantum channel $\mathcal{N}:\mathcal{H}_A \to \mathcal{H}_{A'}$, the coherent information $\mathcal{I}_c(\mathcal{N}, \rho)$ is defined as
\begin{align}
    \mathcal{I}_c(\mathcal{N}, \rho) \equiv S_{A'}-S_{B},
\end{align}
where $S_{A'(B)}$ is the von~Neumann entropy of the output reduced density matrix for subsystem $A'(B)$, and $B$ is an auxiliary system introduced for an isometric embedding of $\mathcal{N}$. In our case, the identification $A=S$, $A'=SM$, and $B=E$ leads to
\begin{align}
\label{seqn:cond_entrpy}
    \mathcal{I}_c(\mathcal{M}, \rho_\textrm{in}) &= S_{SM} - S_E 
    = S_{SM} - S_M 
    = \langle S\rangle,
\end{align}
where the second equality arises from the fact that $\rho_B \simeq \rho_E$ discussed above.
For the third equality, we used the definition of the entropy of the system averaged over different measurement outcomes
\begin{align}
    \langle S \rangle = \sum_i p_i S(\rho_S[i]),
\end{align}
where the index $i$ runs over all possible projective measurement outcomes, $p_i$ is the probability for a particular outcome $i$, and $\rho_S[i]$ is reduced density matrix of the system conditioned on the measurement outcome $i$.
Then, the third equality in Eq.~\eqref{seqn:cond_entrpy} holds because of the block diagonal form of the reduced density matrix for $S$ and $M$~\cite{bao2019theory}:
\begin{align}
    \rho_{SM} = \sum_i p_i \rho_S[i] \otimes \ket{i}_M\bra{i}_M.
\end{align}
Finally, combining Eq.~\eqref{seqn:single_q} and \eqref{seqn:cond_entrpy}, we obtain the key result presented in the main text, which we produce here:
\begin{align}
\label{seqn:relation}
    \mathcal{Q} = \max_{\rho_\textrm{in}} \langle S \rangle.
\end{align}

In general, the optimization over $\rho_\textrm{in}$ for a given channel $\mathcal{M}$ is difficult because it may depends on the detailed information of the unitary evolution as well as set of POVMs.
In the case $\mathcal{M}$ is random, as random circuit models, we can define a closely related quantity to characterize the maximal amount of information that can be transmitted through $\mathcal{M}$ \emph{without a priori specifying the instance $\mathcal{M}$}:
\begin{align}
    \bar{\mathcal{Q}} \equiv \max_{\rho_\textrm{in}} \mathds{E}\left[ \langle S \rangle \right],
\end{align}
where $\mathds{E}$ denotes averaging over random unitary gates and measurement positions.
We note that $\bar{\mathcal{Q}}$ is distinct from average quantum channel capacity $\mathds{E}[\mathcal{Q}]$ since the optimization over $\rho_\textrm{in}$ is performed after averaging over different realization of quantum circuits.
Nevertheless, $\bar{\mathcal{Q}}$ has an operational meaning; it quantifies the maximum amount of coherent information that a randomly chosen random unitary circuit with randomly positioned projective measurements can achieve.
In other words, $\bar{\mathcal{Q}}$ is the capacity for random quantum channel $\mathcal{M}$ in which the random realizations of unitary gates and measurement positions are a priori not known to an encoder.

In Ref.~\cite{gullans2019dynamical}, for a maximally mixed input state $\rho_{\text{in}} = \rho^{\max}$, $\mathds{E}[\langle S \rangle]$ has been identified as an alternative signature of the entanglement phase transition based on numerical simulations of random Clifford circuits with projective measurements.
In more recent work~\cite{bao2019theory}, it has been proven that the transition in the scaling behavior in $\mathds{E}[\langle S \rangle]$ indeed coincides with the entanglement phase transition for Haar random unitary circuits with measurements.
Based on the mapping of such quantum circuits to a class of statistical mechanics models and the replica technique introduced in Ref.~\cite{bao2019theory}, one can show $\mathds{E}[\langle S \rangle ]$ is maximized when $\rho_\textrm{in}$ is maximally mixed.
Therefore, $\bar{\mathcal{Q}}$ exactly corresponds to the the numerical results of $\mathds{E}[\langle S \rangle]$ studied in Ref.~\cite{gullans2019dynamical} and is indeed a signature of the entanglement phase transition.

An alternative way to understand the meaning of $\bar{\mathcal{Q}}$ is to introduce a new channel $\tilde{\mathcal{M}}$ that incorporates the randomness of $\mathcal{M}$ as parts of the definition.
To be more specific, one can consider that unitary gates applied to the system are determined by quantum states of additionally introduced ancilla qudits $M_U$. If the qudits are initialized in equal superposition of their computational basis states, the projective measurements on $M_U$ in the basis realize a certain instance of random unitary gates.
A similar argument can be made for different positioning of measurements on system qudits (with additional ancilla $M_P$).
In this way, the classical information associated with the randomness (in unitary gates and measurement positions) corresponds to the diagonal elements of the density matrix of ancilla $M_{U, P}$.
Following the same analysis described above for $\mathcal{M}$, it follows that $\bar{\mathcal{Q}}$ is the quantum channel capacity of $\tilde{\mathcal{M}}$:
\begin{align}
    \tilde{\mathcal{Q}} = \max_{\rho_{\text{in}}} \mathds{E}\left[\left< S \right> \right] = \bar{\mathcal{Q}}.
\end{align}
This channel capacity is achieved with the maximally mixed input state $\rho_\textrm{in}$ since it must be invariant under any local unitary rotations.
We note that the new channel $\tilde{\mathcal{M}}$ defined by the dynamics of the system and entire collection of ancilla is again a degradable quantum channel following the same reasoning as in the case of $\mathcal{M}$.
This completes the relation between the quantum channel capacity and the averaged entropy of the system conditioned on measurement outcomes.

\section{An improved bound on the entanglement reduction by measurements}\label{sec:decoupling_inequality}

In the previous section, we define a quantum channel $\mathcal{M}$ that describes a generic unitary evolution interspersed by measurements. 
The channel $\mathcal{M}$ acts on both the system and ancilla (representing measurement devices) qubits.
The coherent information $\mathcal{I}_c(\mathcal{M}, \rho_{\text{in}})$ associated with the quantum channel $\mathcal{M}$ describes the transmission of quantum information encoded both in the output state of the system qubits and in the classical measurement outcomes. 
Furthermore, we have shown that the coherent information exactly equals the entanglement entropy of the output quantum state averaged over different classical measurement outcomes [see Eq.~\eqref{seqn:cond_entrpy}].

Motivated by these new understandings, we revisit the analysis of the toy model in Fig.~1 of the main text and present an improved, tight version of the decoupling inequality, analogous to Eq.~(1).
The newly derived decoupling inequality is strictly stronger than the previous one.
In particular, our new inequality predicts that the phase transition point asymptotically approaches to $p_c = 1$ as $m \to \infty$ for the qubit-block model introduced in the main text in the strongly scrambling regime $d/(2m) \gtrsim 1$.
The key idea behind our approach is to explicitly separate out the accessible classical information and inaccessible (lost) information in projective measurement processes.

\begin{figure}[t!]
	\includegraphics[width=\textwidth]{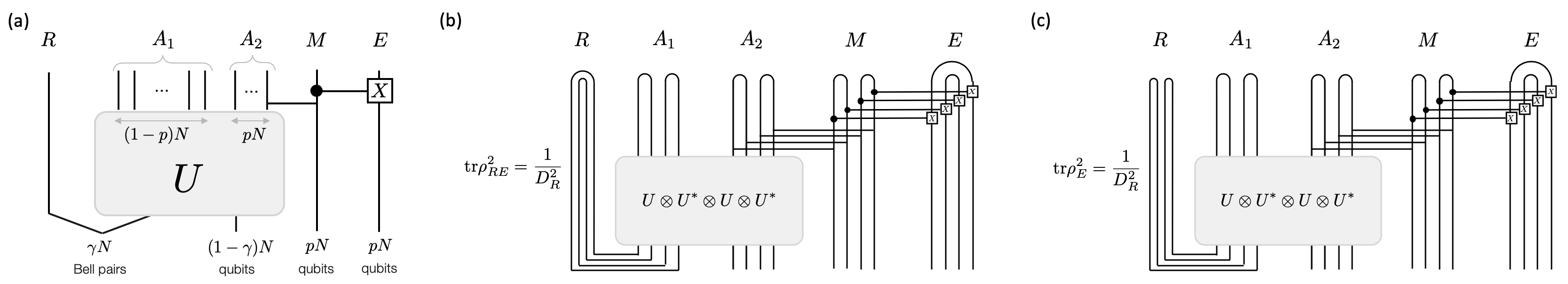}
	\caption{(a) An illustration of the toy model. (b, c) Tensor network representations of $\tr \rho_{RE}^2$ and $\tr \rho_{E}^2$, respectively.}
	\label{suppfig:tn}
\end{figure}

We consider an $N$-qubit system $A$ initially sharing $\gamma N$ bell pairs with the reference $R$.
The rest $(1-\gamma) N$ qubits in the system are prepared in an unentangled product state. 
We apply a random unitary $U \in \textrm{U}(2^N)$ drawn from a unitary $2$-design to the system qubits and perform projective measurements on randomly chosen $pN$ qubits.
We use $A_1$ and $A_2$ to denote the unmeasured and measured qubits, respectively.
Our goal is to show that, under certain conditions, the extensive number of measurements on $A_2$ do not reduce the initial entanglement, $S_R = \gamma N$, between $A$ and $R$.
We show this in two steps: (i) we evaluate the coherent information $\mathcal{I}_c$ of this quantum channel and prove $\mathcal{I}_c \approx S_R = \gamma N$ up to exponentially small corrections in $N$, and (ii) we use Eq.~\eqref{seqn:cond_entrpy} to show that $\langle S \rangle = \mathcal{I}_c = \gamma N$.  
This implies that the average entanglement $\langle S \rangle$ after projective measurements equals its initial value $\gamma  N$, hence no entanglement is reduced by measurements.
Here, we focus on proving (i).

Our first step is to formulate the projective measurements on $A_2$ as weak measurements explained in Sec.~\ref{sec:channel_capacity}.
This formalism is useful to keep track of classical information encoded in measurement outcomes.
Instead of projective measurements, we introduce a set of ancilla qubits $M$ (representing measurement devices), apply entangling unitaries between $A$ and $M$ (to represent the measurement processes), and then dephase $M$ with a dephasing bath $E$ by applying control-$X$ gates between $M$ and $E$.
The last dephasing procedure is crucial and necessary because a measurement device can hold only classical information (i.e., diagonal elements of its reduced density matrix in computational basis).
Using this formalism does not lead to any loss of generality.
An illustration of the setup is presented in Fig.~\ref{suppfig:tn}(a).
We are interested in how much of the initial entanglement ($\gamma N$ Bell pairs) remains (can be recovered) in the quantum state of $A$ \emph{and} classical measurement outcomes in $M$. This can be quantified by the coherent quantum information~\cite{devetak2005capacity}:
\begin{align}
    \mathcal{I}_c(\mathcal{M}, \rho_{\text{in}}) \equiv S_R - I(R:E),
\end{align}
which implies that the reduction in $\mathcal{I}_c$ is determined by the mutual information $I(R:E)$ between the reference and environment.
Therefore, it suffices to show $I(R:E) = 0$.
Instead of directly evaluating the vanishing mutual information, we show an equivalent statement that the reference $R$ and the dephasing bath $E$ are decoupled,
\begin{align}
    \label{seqn:new_decoupling}
    \rho_{RE} \approx \rho_R \otimes \rho_E
\end{align}
up to exponentially small corrections, 
provided that inequality $p < 1 - \gamma$ is satisfied.
More specifically, we provide an upper bound to the $L_1$-distance between both sides of Eq.~\eqref{seqn:new_decoupling}.
According to the Cauchy-Schwarz inequality, we have
\begin{align}
\lVert \rho_{RE} - \rho_R \otimes \rho_E \lVert_1^2 &\leq D_R D_E \tr \left[ \left( \rho_{RE} - \rho_{R} \otimes \rho_E\right)^2 \right] \\
    \label{seqn:CS_inequality}
&= 2^{\gamma N} 2^{pN} \left[ \tr \rho_{RE}^2 - \frac{1}{2^{\gamma N}} \tr \rho_E^2 \right],
\end{align}
where $D_R = 2^{\gamma N}$ and $D_E = 2^{p N}$ are the Hilbert space dimension of the reference $R$ and dephasing bath $E$, respectively, and the reduced density matrix for the reference is maximally mixed, i.e., $\rho_R = 2^{-\gamma N} \mathds{1}_R $. 
Two terms,  $\tr \rho_{RE}^2$ and $\tr \rho_E^2$ on the right-hand side, can be written as the expectation value of a swap operator in a duplicated Hilbert space~\cite{PreskillNotes}: 
\begin{align}
    \label{seqn:swap_trick}
    \tr \rho_X^2 = \tr \left[ \textrm{SWAP} \left(\rho_X \otimes \rho_X\right) \right] \;\;\; \textrm{with} \;\;\;  X\in \{\textrm{RE}, \textrm{E} \},
\end{align}
where $\textrm{SWAP}$ is an operator that swaps wavefunctions in two different copies. 
Equation~\eqref{seqn:swap_trick} can be pictorially represented by using tensor network diagrams in Figs.~\ref{suppfig:tn}(b,c), where taking the trace or the expectation values of $\textrm{SWAP}$ correspond to different contractions at the top of diagrams.

\begin{figure}[t!]
	\includegraphics[width=0.7\textwidth]{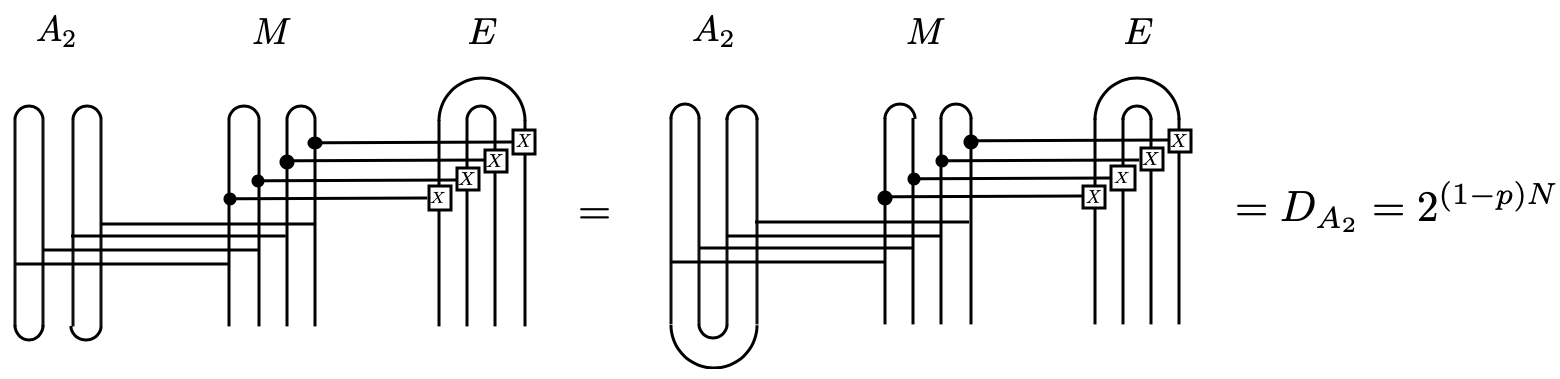}
	\caption{
	Tensor network diagrams associated with the measured qubits $A_2$, measurement device $M$ and the dephasing bath $E$ in Eqs.~\ref{eqn:contraction1} and \ref{eqn:contraction2}.}
	\label{suppfig:calc}
\end{figure}

Evaluating the right hand side of Eq.~\eqref{seqn:CS_inequality} for an arbitrary $U$ is computationally intractable.
However, we can exactly evaluate it once it is averaged over all possible unitaries in $\textrm{U}(2^N)$ in Haar measure (or any unitary $2$-design).
On the right-hand side, evaluating the average $\tr\rho_{RE}^2$ and $\tr\rho_E^2$ involves computing the second moment $\mathbb{E}_U[U\otimes U^* \otimes U \otimes U^*]$ of random unitary $U$:
\begin{align}
\mathbb{E}_U\left[{U\otimes U^* \otimes U \otimes U^*}\right]
=
\sum_{\sigma,\tau =\pm1} w_g^{(2)}(\sigma,\tau) \hat{\tau}_{\mathbf{ab}}
\hat{\sigma}_{\mathbf{cd}}, \label{eqn:2-design_avg}
\end{align}
where the coefficient
\begin{align}
\label{eqn:weingarten_coeff}
w_g^{(2)}(\sigma,\tau) = 
\frac{\delta_{\sigma,\tau}}{d^2-1}  
-\frac{1-\delta_{\sigma,\tau}}{d(d^2-1)},
\end{align}
which is the so-called Weingarten function~\cite{collins2003moments}, and $\hat{\sigma}$ and $\hat{\tau}$
are tensors associated with the binary variables $\sigma, \tau \in \{\pm 1\}$ defined as 
\begin{align}
\label{eqn:tensor_content}
    \hat{\xi}_{\mathbf{ab}} 
    =\left\{
\begin{array}{ll}
\delta_{a_1b_1} \delta_{a_2b_2} & \textrm{if  } \xi = +1\\
\delta_{a_1b_2} \delta_{a_2b_1} & \textrm{if  } \xi = -1
\end{array}
\right.
.
\end{align}
Using this property, we can explicitly evaluate $\mathbb{E}_U\left[ \tr \rho_{RE}^2 \right]$ and $\mathbb{E}_U\left[ \tr \rho_{E}^2 \right]$:
\begin{align}
\mathbb{E}_U\left[ \tr \rho_{RE}^2 \right] =& \frac{1}{D_R^2} \bigg[ \frac{1}{D_A^2 - 1} D_{R} D_{A_1}^2  - \frac{1}{D_A(D_A^2 - 1)} D_R^2 D_{A_1}^2 - \frac{1}{D_A(D_A^2 - 1)} D_R D_{A_1} + \frac{1}{D_A^2 - 1} D_{R}^2 D_{A_1} \bigg] D_{A_2} \nonumber \\
\simeq& 2^{-(\gamma+p)N} - 2^{-(1+p)N} - 2^{-(2+\gamma)N} + 2^{-N}, \label{eqn:contraction1}\\
\frac{1}{2^{\gamma N}}\mathbb{E}_U\left[ \tr \rho_{E}^2 \right] =& \frac{1}{2^{\gamma N}} \frac{1}{D_R^2} \bigg[ \frac{1}{D_A^2 - 1} D_{R}^2 D_{A_1}^2  - \frac{1}{D_A(D_A^2 - 1)} D_R D_{A_1}^2 - \frac{1}{D_A(D_A^2 - 1)} D_R^2 D_{A_1} + \frac{1}{D_A^2 - 1} D_{R} D_{A_1} \bigg] D_{A_2} \nonumber \\
\simeq& 2^{-(\gamma+p)N} - 2^{-(1+2\gamma+p)N} - 2^{-(2+\gamma)N} + 2^{-(1+2\gamma)N}, \label{eqn:contraction2}
\end{align}
where $D_{A_1} = 2^{(1-p)N}$, $D_{A_2} = 2^{pN}$ and $D_A = 2^N$ are the Hilbert space dimension of subsystem $A_1$, $A_2$ and system $A$, respectively, and we considered the limit $N \gg 1$ in the second line of both equations.
In the first equalities of Eqs.~\ref{eqn:contraction1} and~\ref{eqn:contraction2}, we used the results of tensor contractions presented in Fig.~\ref{suppfig:calc}.
Plugging above two results into Eq.~\eqref{seqn:CS_inequality}, we obtain a new decopuling inequality. 
In particular, in the limit $N \gg 1$, the decoupling inequality becomes
\begin{align}
\mathbb{E}_U\left[ \lVert \rho_{RE} - \rho_R \otimes \rho_E \lVert_1 \right] \leq \sqrt{\mathbb{E}_U\left[ \lVert \rho_{RE} - \rho_R \otimes \rho_E \lVert_1^2 \right]} \lesssim 2^{-(1-\gamma - p) N/2}.
\end{align}

We find that when $p < 1 - \gamma$ with $N\rightarrow \infty$, $\rho_{RE}$ factorizes as desired.
We emphasize that this condition is strictly stronger than the n\"aive sufficient condition $p < (1 - \gamma)/2$ presented in the main text, which demands the decoupling between $R$ and $A_2$.
At technical level, the distinction between two results arises from the fact that we include dephased measurement devices $M$ as part of accessible information in the present approach.

Using this newly derived inequality, we can obtain quantitative prediction on the critical measurement probability in the $1$D qubit-block model. When $d/(2m) \gtrsim 1$, we can approximate the local Clifford circuit by a random unitary gate drawn from a unitary 2-design acting on $2m$ qubits. Further considering the limit $m \gg 1$, we can directly apply our analysis here to estimate the amount of entanglement reduction due to measurements.
The decoupling inequality above suggests no (or exponentially small in $m$) entanglement reduction as long as the entanglement entropy per qubit cluster $\gamma < 1 - p$. 
Since the phase transition occurs when the entanglement density vanishes $\gamma = 0$, we expect $p_c = 1$, provided $d/(2m) \gtrsim 1$ and $m \rightarrow \infty $.
This prediction suggests $p_c(m)$ in Fig.~\ref{suppfig:transition_m}(a) should asymptotically approach unity in the strongly scrambling regime $d/(2m) \gtrsim 1$ as $m \to \infty$, consistent with our numerical simulation results.

\bibliography{refs}